\documentclass[12pt]{iopart}
\usepackage{graphicx}
\usepackage{booktabs}
\usepackage{amssymb}
\usepackage{amsfonts}
\usepackage[justification=centering]{caption}
\usepackage{float}
\usepackage[printwatermark]{xwatermark}
\usepackage{xcolor}
\usepackage{textcomp}
\usepackage{epstopdf}
\usepackage{gensymb}
\usepackage{blindtext}
\usepackage{siunitx}
\usepackage{tabularx}
\usepackage{enumitem}
\usepackage{subfig}
\usepackage{fourier} 
\usepackage{makecell}
\expandafter\let\csname equation*\endcsname\relax
\expandafter\let\csname endequation*\endcsname\relax
\usepackage{amsmath}

\newcommand{\code}[1]{\texttt{#1}}

\begin{document}
\title{Beyond the Limits of 1D Coherent Synchrotron Radiation}%
\author{A.~D.~Brynes\textsuperscript{1,2,3}, P.~Smorenburg\textsuperscript{4}, I.~Akkermans\textsuperscript{4}, E.~Allaria\textsuperscript{5}, L.~Badano\textsuperscript{5}, S.~Brussaard\textsuperscript{4}, M.~Danailov\textsuperscript{5}, A.~Demidovich\textsuperscript{5}, G.~De~Ninno\textsuperscript{5}, D.~Gauthier\textsuperscript{5,6}, G.~Gaio\textsuperscript{5}, S.~B.~van~der~Geer\textsuperscript{7}, L.~Giannessi\textsuperscript{5}, M.~J.~de~Loos\textsuperscript{7}, N.~S.~Mirian\textsuperscript{5}, G.~Penco\textsuperscript{5}, P.~Rebernik\textsuperscript{5}, F.~Rossi\textsuperscript{5}, I.~Setija\textsuperscript{4}, S.~Spampinati\textsuperscript{5}, C.~Spezzani\textsuperscript{5}, M.~Trov\`o\textsuperscript{5}, P.~H.~Williams\textsuperscript{1,2} \& S.~Di~Mitri\textsuperscript{5}}
\address{\textsuperscript{1} ASTeC, STFC Daresbury Laboratory, Daresbury, Warrington, WA4 4AD Cheshire, United Kingdom}
\address{\textsuperscript{2} Cockcroft Institute, Sci-Tech Daresbury, Keckwick Lane, Daresbury, Warrington WA4 4AD, United Kingdom}
\address{\textsuperscript{3} Department of Physics, University of Liverpool, Liverpool, L69 7ZE, United Kingdom}
\address{\textsuperscript{4} ASML Netherlands B.V., 5504 DR Veldhoven, Netherlands}
\address{\textsuperscript{5} Elettra-Sincrotrone Trieste S.C.p.A., 34149 Basovizza, Trieste, Italy}
\address{\textsuperscript{6} LIDYL, CEA, CNRS, Universit\'e Paris-Saclay, Saclay, 91191 Gif-sur-Yvette, France}
\address{\textsuperscript{7} Pulsar Physics, Burghstraat 47, 5614 BC Eindhoven, Netherlands}

\begin{abstract} 
	An understanding of collective effects is of fundamental importance for the design and optimisation of the performance of modern accelerators. In particular, the design of an accelerator with strict requirements on the beam quality, such as a free electron laser (FEL), is highly dependent on a correspondence between simulation, theory and experiments in order to correctly account for the effect of coherent synchrotron radiation (CSR), and other collective effects. A traditional approach in accelerator simulation codes is to utilise an analytic one-dimensional approximation to the CSR force. We present an extension of the 1D CSR theory in order to correctly account for the CSR force at the entrance and exit of a bending magnet. A limited range of applicability to this solution -- in particular, in bunches with a large transverse spot size or offset from the nominal axis -- is recognised. More recently developed codes calculate the CSR effect in dispersive regions directly from the Li\'enard-Wiechert potentials, albeit with approximations to improve the computational time. A new module of the General Particle Tracer (\textsc{GPT}) code was developed for simulating the effects of CSR, and benchmarked against other codes. We experimentally demonstrate departure from the commonly used 1D CSR theory for more extreme bunch length compression scenarios at the FERMI FEL facility. Better agreement is found between experimental data and the codes which account for the transverse extent of the bunch, particularly in more extreme compression scenarios.
\end{abstract}

\maketitle

\section{Introduction}

The emission of coherent synchrotron radiation (CSR) on curved trajectories can present a significant issue for short electron bunches, such as those used in free electron lasers (FELs) \cite{PhysRep.539.1,PhysRevAccelBeams.19.034402,PhysRevSTAB.18.030706,PhysRevSTAB.12.030704,PhysRevSTAB.15.020701}. CSR can degrade the quality of electron bunches through an increase in projected and slice emittance, and energy spread. As FEL facilities push for more exotic lasing schemes, and the requirements for drive bunches become more stringent, there is an increasing demand for accurate techniques to both measure and simulate the bunch properties throughout the accelerator. 

A number of codes exist which are capable of simulating the effects of CSR, some of which utilise a 1D approximation, based on Ref.\,\cite{NIMA.1997.2.373}, and others which extend the model to incorporate 2D and 3D effects. While previous studies have shown good agreement between results from some of these simulation codes and experimental data \cite{PhysRevSTAB.18.030706,PhysRevSTAB.12.030704}, there is a point at which the 1D approximation is no longer valid, as given by the Derbenev criterion \cite{TESLA-FEL-Report-1995-05}, which suggests that projecting the bunch distribution onto a line may overestimate the level of coherent emission, particularly when the bunch has a large transverse-to-longitudinal aspect ratio. The primary aim of this study is to determine if, during strong bunch compression, or for bunches with a large transverse-to-longitudinal aspect ratio, the limits of the 1D approximation could be found. This is achieved through comparing analytic results with simulation codes that incorporate the transverse bunch distribution, and with experimental data. The projected emittance of the electron beam was measured in parameter scans at the exit of the first bunch length compressor of the FERMI FEL \cite{NatPhoton.6.699,NatPhoton.7.913}. This benchmarking study of CSR is accompanied by new insights on the CSR transient field at the edges of dipole magnets, which suggest novel compressor designs for the minimization of this instability. A new CSR feature of the General Particle Tracer (\textsc{GPT}) \cite{GPT} tracking code was developed specifically for this study.

The paper is organised as follows. In Sec.\,\ref{sec:csrcalculation} we derive the longitudinal CSR force in three regimes: the entrance transient regime, in which the entire bunch has not yet entered the magnet; the steady-state regime, when the entire bunch is travelling through the magnet; and the exit transient regime, at which point the head of the bunch has left the dipole, but the tail is still radiating. In Sec.\,\ref{sec:numericalvalidation} a numerical simulation of the 1D CSR force is performed and the results are compared with analytical predictions. A further examination of the impact of the transverse extent of the bunch with respect to the CSR force is given in Sec.\,\ref{sec:transitionregime}, demonstrating the issue with projecting the force entirely onto the longitudinal dimension. The FERMI facility is briefly outlined in Sec.\,\ref{sec:parameterscans} along with details of the parameter scans undertaken to measure the projected emittance of the bunch as a function of compression, longitudinal distribution and matching. A comparison of the codes used for validating the simulation of CSR is given in Sec.\,\ref{sec:simulationsetup}, and a comparison between theory, simulation and experiment is discussed in Sec.\,\ref{sec:results}. Finally, we summarise our findings in Sec.\,\ref{sec:conclusions}.

\section{Calculation of CSR force} \label{sec:csrcalculation}

This section will provide an extension of the 1D CSR force first calculated in \cite{NIMA.1997.2.373}, and subsequently expanded on in \cite{SLAC-PUB-9242}, by deriving new expressions for the CSR force at the entrance and exit of a bending magnet. We first consider the situation of a bunch of electrons on a curved trajectory at time $t$ through a bending magnet of bending radius $R$ and bending angle $\phi_m$. The electromagnetic field acting upon any particular electron in the bunch is comprised of the fields emitted by electrons at earlier times $t' < t$ on this curved path. In the following derivation, the subscripts $0$ and $1$ will refer to the emitting and receiving particle, respectively, and a prime indicates retarded time or position; that is, the point at which the field was emitted. To calculate the total field, we first consider the field emitted by a single electron at position $\vec{r_0}'$  inside the magnet at time $t'$ and observed by another electron at position $\vec{r_1}$ at time $t$. For simplicity, in this section we neglect the transverse extent and energy spread of the electron bunch, and thereby assume that all electrons travel exactly along the reference trajectory. The electromagnetic field at $\vec{r_1}$ due to the electron at $\vec{r_0}'$ is given by the well-known Li\'enard-Wiechert field \cite{Jackson} at time $t$:

\begin{equation} \label{eq:lwfield}
\vec{E}(\vec{r},t) = \frac{e}{4 \pi \epsilon_0}\left(\frac{\vec{n} - \vec{\beta}'}{\gamma^2 (1 - \vec{n} \cdot \vec{\beta}')^3 \rho^2} + \frac{\vec{n}\times \left(\left(\vec{n} - \vec{\beta}'\right) \times \dot{\vec{\beta}}'\right)}{c \left(1 - \vec{n}\cdot \vec{\beta}'\right)^3 \rho}\right),
\end{equation}

\noindent where $e$ is the electron charge, $\epsilon_0$ is the vacuum permittivity, $c$ is the speed of light, $\gamma$ is the relativistic Lorentz factor, $\vec{\beta_0}'$ is the velocity of the emitting electron (normalised to $c$), $\dot{\vec{\beta_0}}'$ is the normalised acceleration of the emitting electron, $\rho = |\vec{r_1}-\vec{r_0}'|$ is the distance between the emission site and point of observation, and $\vec{n}=(\vec{r_1}-\vec{r_0}')/\rho$. From now on, the first term of Eq.\,\ref{eq:lwfield}, which does not depend on $\dot{\vec{\beta_0}}'$, will be referred to as the \lq velocity\rq\,or \lq Coulomb\rq\,field, and we will refer to the second term as the \lq radiation\rq\,field. Conventionally, several regimes of CSR forces are identified according to the positions of the emitting and observing particles \cite{NIMA.1997.2.373}. Initially, the particle in front is inside the magnetic field of the dipole and the particle behind has not yet entered it, in which case $\dot{\vec{\beta_0}} = 0$ and only the first term of Eq.\,\ref{eq:lwfield} contributes, known as the \lq entrance transient\rq\,regime. When both particles are inside the magnet, both terms in Eq.\,\ref{eq:lwfield} contribute to the CSR field, and this is known as the \lq steady-state\rq\,regime. Finally, when the emitter is still in the magnet and the receiver has exited it, this is known as the \lq exit transient\rq\,regime. Eq.\,\ref{eq:lwfield} describes the electric field due to a single point particle, and so to calculate the entire CSR field requires a convolution of this expression with the charge density of the entire bunch, using the general expression for the longitudinal CSR wake $E_{||}$ at a given position $z$:

\begin{equation}\label{eq:generalcsrfield}
E_{||} (z) = N e \int w(z - z') \lambda(z') dz',
\end{equation}

\noindent where $N$ is the number of particles in the bunch, $e$ the electron charge, $\lambda(z)$ is the longitudinal charge distribution, with the normalisation condition $\int \lambda(s) ds = 1$, and $w(z - z')$ is the parallel component of the field in Eq.\,\ref{eq:lwfield} at position $z$ in the bunch due to a particle at position $z'$ in the bunch. 

\subsection{Steady-State Regime}

\begin{figure}[h]
	\begin{center}
		\centering           
		\includegraphics[width=8cm]{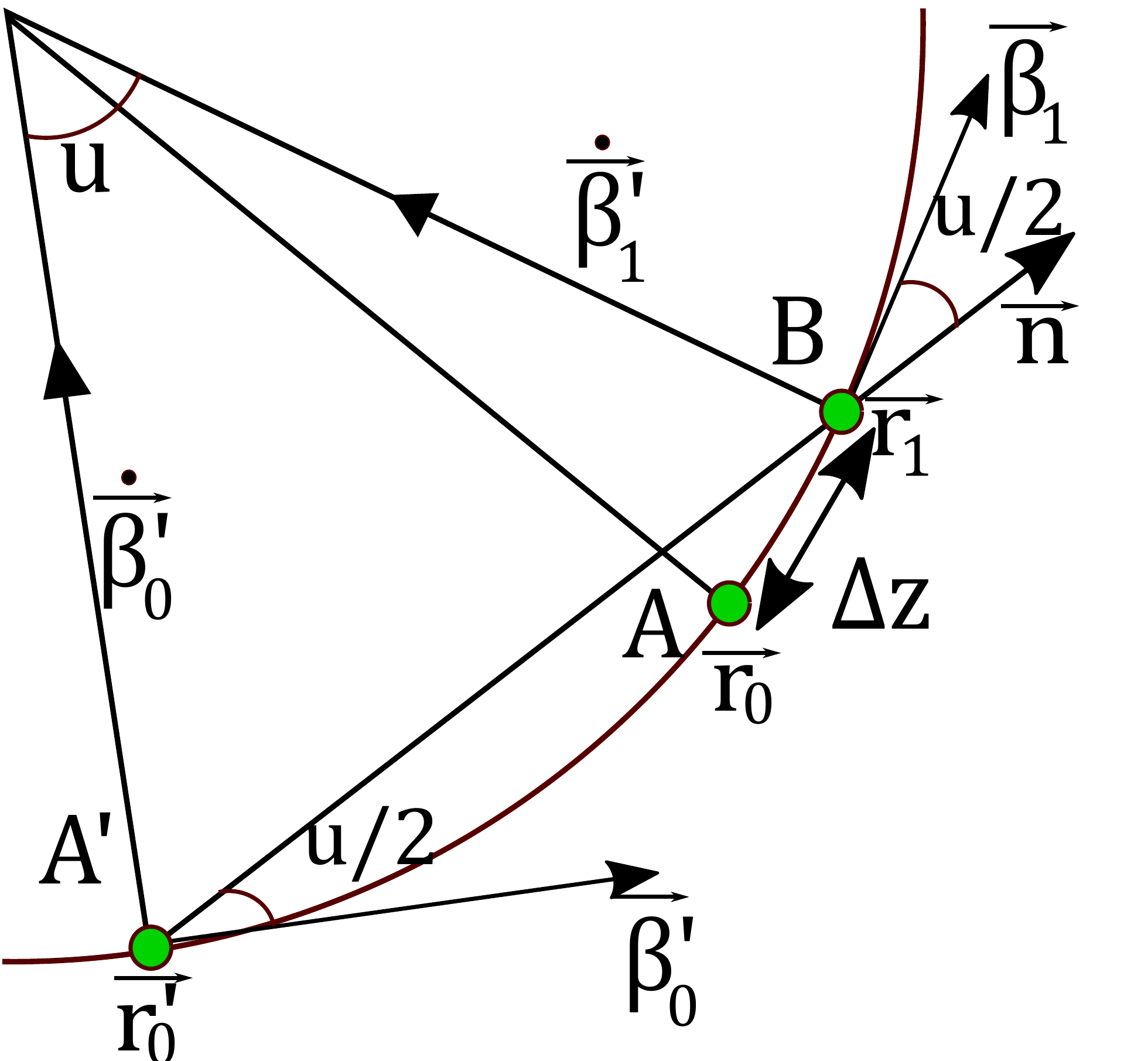}
		\caption{Schematic of the CSR interaction in the 1D model in the steady-state regime. The positions of the emitter and receiver at the time of interaction are shown as $A$ and $B$, the position of the emitter at the time of emission is $A'$.}
		\label{fig:steady_state_csr}
	\end{center}
\end{figure}

As shown in \cite{NIMA.1997.2.373,TESLA-FEL-Report-1995-05}, the electric field observed at position $z$ for a line charge $\lambda(z)$ due to the motion on a circular arc of radius $R$ is given by:

\begin{equation}\label{eq:steadystateapprox}
E_{||}^{SS}(z) = \frac{N e \beta^2}{8 \pi \epsilon_0 R} \int_{0}^{\phi} \frac{\beta - \cos(u/2)}{\left(1 - \beta \cos(u/2)\right)^2} \lambda(z - \Delta z(u))du,
\end{equation}

\noindent where $\phi$ is the angle from the entrance of the magnet to the observation point, $\Delta z(u) = R(u - 2\beta \sin(u/2))$ and $u$ is the retarded angle between the emitter at time of emission and the observer at time of observation. A schematic of this scenario is shown in Fig.\,\ref{fig:steady_state_csr}. Note that both the position of the emitting electron at time of emission $\vec{r_0}'$ and at time of observing $\vec{r_0}$ have been drawn, reflecting the fact that the bunch travels a considerable distance during the time required for the electromagnetic field to travel from emitter to observer. This formula is valid for a rigid line charge, using the ultrarelativistic approximation ($\beta \approx 1$). This model also does not take account of any effects due to dipole fringe fields. The transition to the steady-state regime takes place at a distance $D^{SS}$ from the entrance to the magnet \cite{NIMA.1997.2.373}:

\begin{equation}\label{eq:steadystatecondition}
D^{SS} \approx \left(24 R^2 \sigma_z \right)^{1/3},
\end{equation}

\noindent with $\sigma_z$ the rms bunch length.

\subsection{Entrance Transient Regime}\label{sec:entrancetransient}

\begin{figure}
\begin{center}
	\centering           
	\includegraphics[width=8cm]{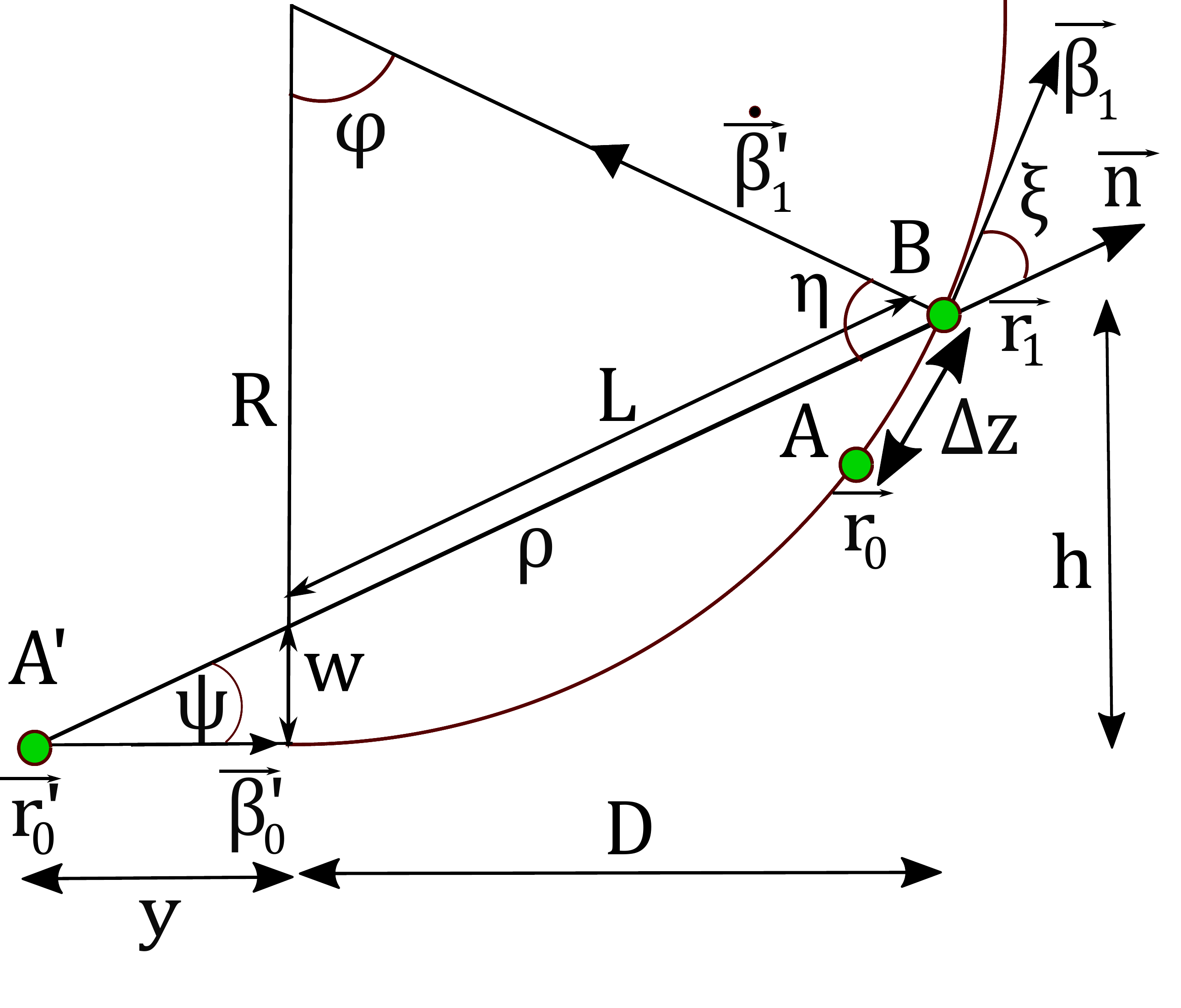}
	\caption{Geometry of CSR interaction between an emitting electron at $\vec{r_0}'$ before the magnet and a receiving electron at $\vec{r_1}$ within the magnet. $\psi$ is the angle between the emitter at time $t'$ and the receiver at time $t$, and $\xi$ is the angle between $\vec{\beta}$ and $\vec{n}$. The distances and angles shown in this schematic are derived in \ref{sec:appendix_entrance}.}
	\label{fig:csr_entrance_transient}
\end{center}
\end{figure}

In this regime, the condition $D^{SS}$ has not been reached, and a significant portion of the emitting particles have not yet entered the magnetic field. This means that their contribution comes entirely from the velocity field of Eq.\,\ref{eq:lwfield}. For the full derivation of the total CSR field in this regime, see \ref{sec:appendix_entrance}. The resulting expression for this field is:

\begin{equation}\label{eq:entrancetransientfield}
E_{||}^{ent}(z) = E_{||}^{SS}(z) + \frac{Ne}{4\pi\epsilon_0 \gamma^2}\int_{0}^{d}\frac{(y - \beta \rho(y))\cos(\phi) + R\sin(\phi)}{(\rho(y) - \beta(y + r\sin(\phi)))^2 \rho(y)} \lambda(z - \Delta(y))dy,
\end{equation}

\noindent where $\rho(y) = \sqrt{y^2 + 2Ry\sin(\phi) + 4R^2\sin^2 (\phi/2)}$ and $\Delta(y) = y + R\phi - \beta\rho(y)$, with $y$ the distance between the emitting particle and the entrance of the magnet, and $d$ the length of the drift before the magnet taken into account for the calculation of the CSR field. A representation of this regime is shown in Fig.\,\ref{fig:csr_entrance_transient}. The contributions to the field from $E_{||}^{SS}$ arise from the radiative emission of particles on the curved trajectory, while the other term comes from particles which have not yet reached the magnet at the time of emission. Both terms of Eq.\ref{eq:entrancetransientfield} partially cancel, and the net CSR field has a lower amplitude than either term. In the limit of the drift before the bend $d \to \infty$, in the small-angle and ultrarelativistic approximations \cite{NIMA.1997.2.373}, this field reduces to:

\begin{equation}\label{eq:steadystatefield}
E_{||}^{ent}(z) = \frac{e}{24^{1/3}\pi \epsilon_0 R^{2/3}} \left( \left(\frac{24}{R\phi^3}\right)^{1/3} \left[\lambda \left(z - \frac{R\phi^3}{24}\right) - \lambda \left(z - \frac{R\phi^3}{6}\right)\right] + \int_{z - R\phi^3/24}^{z} \frac{d\lambda(z')}{dz'}\frac{dz'}{\left(z - z'\right)^{1/3}}\right).
\end{equation}

\noindent However, without taking this limit, that is, if $d$ is small, the contribution from the velocity term is smaller than expected from Eq.\,\ref{eq:steadystatefield} and the radiation term dominates.

\begin{figure}
\centering
\subfloat[Although Coulomb field lines align with the current position, the field actually originates from the retarded position.]{  
	\includegraphics[width=5cm]{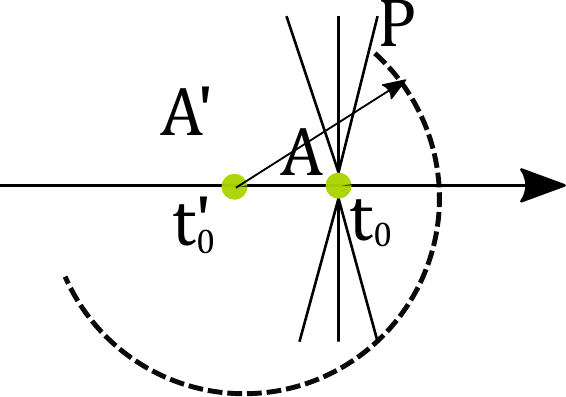} 
	\label{fig:entrance_case_a}}
\hfill
\subfloat[Conseqeuently, if the emitting electron $A'$ electron bends between the retarded time $t_0'$ and the current time $t_0$, the field at the observation point is not affected.]{
	\includegraphics[width=5cm]{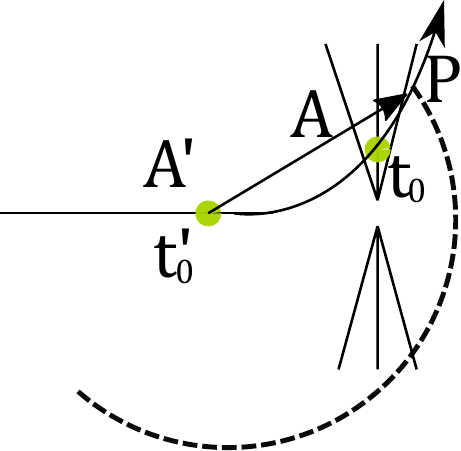}
	\label{fig:entrance_case_b}}

\subfloat[This allows electrons ($B$) in front of the source electron ($A'$) to end up inside the high-field region due to $A'$.]{
	\includegraphics[width=5cm]{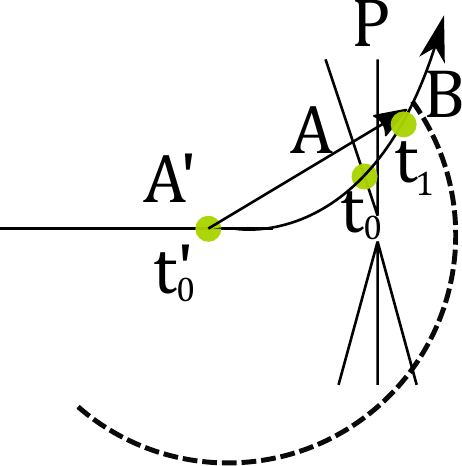}
	\label{fig:entrance_case_c}}
\hfill
\subfloat[Whether the high-field region exists at time $t_1$ depends on the length of the drift on which the source electron has been moving.]{
	\includegraphics[width=5cm]{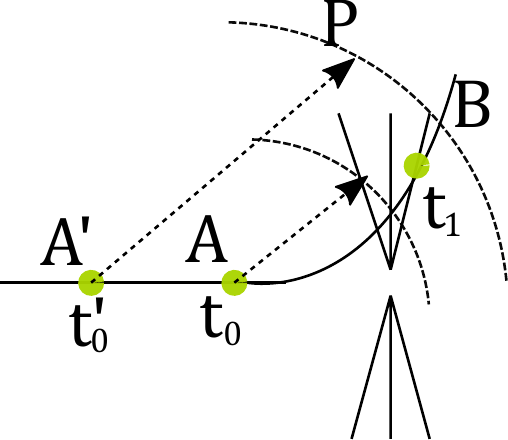} 
	\label{fig:entrance_case_d}}

\caption{Schematic representation of the CSR entrance transient field. The observation point $P$ due to the Li\'enard-Wiechert field of the emitter $A'$ at a retarded time $t_0'$ is shown as the dashed circle. }
\label{fig:entrance_transient_schematic}
\end{figure}

This result -- that the velocity component of the Li\'enard-Wiechert field can provide a non-negligible contribution to the CSR field in the entrance transient regime, even in the ultrarelativistic limit -- can be understood in the following way, as illustrated by Fig.\,\ref{fig:entrance_transient_schematic}. The Coulomb field of a particle on a straight trajectory is confined to a narrow disk, and it appears to be produced instantaneously by the electron at position $r_0$, at time $t_0$ to an observer, whereas in fact the field was produced at a retarded time $t_0'$ (Fig.\,\ref{fig:entrance_case_a}). Even if the electron subsequently moves onto a different trajectory between $t_0'$ and $t_0$, this will not change the field at the observation point, and the Coulomb field is still travelling along the straight path (Fig.\,\ref{fig:entrance_case_b}). For a bunch of electrons beginning to enter a curved path, the electrons at the head will observe this Coulomb field generated by the tail of the bunch, as the field from the tail has been able to \lq catch up\rq\, with the head, which has taken a longer time to travel the same longitudinal distance along the initial axis (Fig.\,\ref{fig:entrance_case_c}). This model suggests that, for a given angle $\phi$ into the magnet, there exists a characteristic drift length $d_c$ needed to generate Coulomb fields at that position -- that is, to have an entrance transient effect (Fig.\,\ref{fig:entrance_case_d}). This distance can be estimated by calculating the required distance in front of the magnet that the field would need in order to be observed by the observing electron, giving:

\begin{equation}\label{eq:entrancetransientcatchup}
d_c \approx \frac{\gamma R \phi^2}{\sqrt{2}}.
\end{equation}

\noindent Depending on the length of the drift section preceding the bend, the effect of the entrance transient will have a varying effect, and so it is necessary to take $d_c$ into account in order to correctly account for this. For our benchmark case (see Sec.\,\ref{sec:numericalvalidation}), and taking $\phi \approx (24 \sigma_z / R)^{1/3}$ from Eq.\,\ref{eq:steadystatecondition}, this required distance is $d_c \approx \gamma R^{1/3} \sigma_z^{2/3} \approx 8$\,\si{\metre}. Distances of the order of tens of metres can be incorporated into simulations of bunch compressors for linear machines, but this cannot be done for bends in circular accelerators due to the higher concentration of dipoles. This means that errors can be made if the formula Eq.\,\ref{eq:steadystatefield} is applied in these scenarios, or if a sufficient drift is not taken into account before the entrance to a dispersive region.

\subsection{Exit Transient Regime}\label{sec:exittransient}

\begin{figure}
	\begin{center}
		\centering           
		\includegraphics[width=6.5cm]{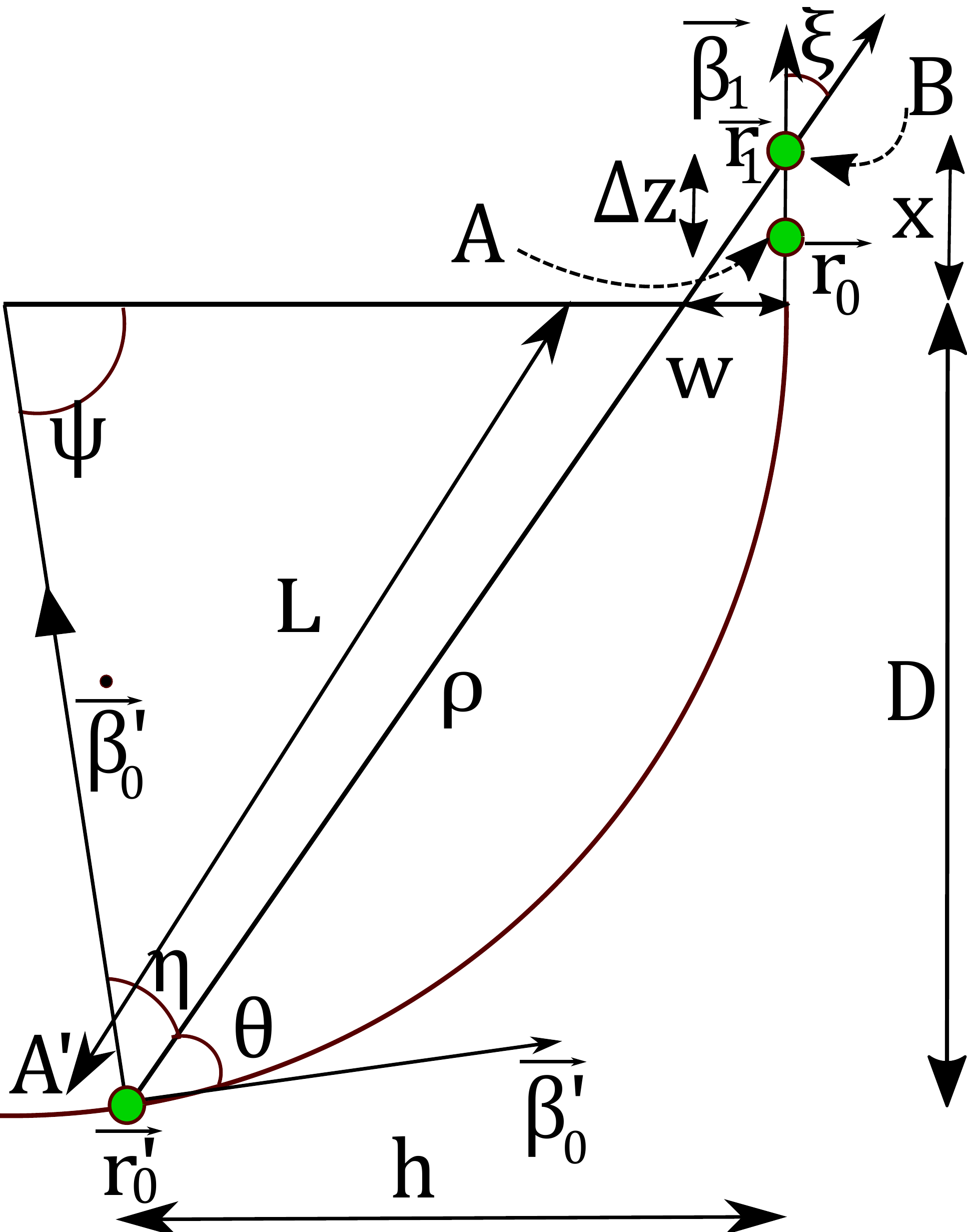}
		\caption{Geometry of CSR interaction between an emitting electron at $\vec{r_0}'$ inside the magnet and a receiving electron at $\vec{r_1}$ past the magnet. The distances and angles shown are derived in \ref{sec:appendix_exit}.}
		\label{fig:csr_exit_transient}
	\end{center}
\end{figure}

Consider the situation at time $t$ which an electron bunch has traveled through a bending magnet of bending radius $R$ and bending angle $\phi_m$ and is currently a distance $x_c$ past the exit edge of the magnet. The electromagnetic field acting upon any particular electron in the bunch  is comprised of the fields emitted by electrons at earlier times $t' < t$ when they were still inside the magnet. To calculate the total field, we first consider the field emitted by a single electron at position $\vec{r_0}'$  inside the magnet at time $t'$ and observed by another electron at position $\vec{r_1}$ past the magnet at time $t$. The geometry of this case is sketched in Fig.\,\ref{fig:csr_exit_transient}.

For the full derivation of the total CSR field after the exit of the bending magnet, see \ref{sec:appendix_exit}. Defining the following quantities:

\begin{equation}
\zeta = x + R\sin(\psi) - \beta \rho \cos(\psi),
\end{equation}

\begin{equation}
\chi = R\sin(\psi/2) + x\cos(\psi/2),
\end{equation}

\noindent with $\psi$ the angle between the emitting electron at retarded position $\vec{r_0}'$ and the exit of the magnet, we obtain the following expression for the radiation field:

\begin{equation} \label{eq:totalcsrfieldexit}
E_{||, rad}^{exit}(z, x) = \frac{Ne \beta^2}{4\pi \epsilon_0}\int_{0}^{\phi_m}\left( \frac{2\sin(\psi/2)\zeta \chi}{\left(\rho - \beta\left(R\sin(\psi) + x\cos(\psi)\right)\right)^2\rho} - \frac{\sin(\psi)}{\rho - \beta\left(R\sin(\psi) + x\cos(\psi)\right)}\right) \lambda\left(z'(\psi)\right)d\psi.
\end{equation}

In this expression, $x = x_c + z$ is the position of the evaluation point with respect to the exit edge of the magnet, with $x_c$ the distance from exit edge to bunch centroid and $z$ the position relative to the bunch centroid. In the integral, the charge density should be evaluated at $z'$, which from Eq.\,\ref{eq:retardationcondition1} is given by $z'(\psi) = -x_c - R\psi + \beta\rho$. The corresponding expression for the velocity field is:

\begin{equation} \label{eq:totalcsrfieldvelocityexit}
E_{||, vel}^{exit}(z, x) = \frac{Ne \beta R}{4\pi \epsilon_0 \gamma^2}\int_{0}^{\phi_m} \frac{x - \beta \rho \cos(\psi) + R \sin(\psi)}{\left(\rho - \beta\left(R \sin(\psi) + x \cos(\psi)\right)\right)^2 \rho} \times \lambda(z'(\psi))d\psi.
\end{equation}

\noindent Eq.\,\ref{eq:totalcsrfieldexit} gives the longitudinal radiation field as observed along a bunch that just passed a single bending magnet. The expression for the radiation field (Eq.\,\ref{eq:totalcsrfieldexit}) can be integrated using the ultrarelativistic and small-angle approximations (see \ref{sec:appendix_exit}) to yield the full field:

\begin{equation} \label{eq:totalcsrexitwakeradiation}
E_{||, rad}^{exit}(z, x) \approx \frac{Ne}{\pi \epsilon_0}\left(\frac{\lambda (z - \Delta z_{max})}{\phi_m R + 2x} - \frac{\lambda(z)}{2x} + \int_{z - \Delta z_{max}}^{z} \frac{\partial \lambda (z')}{\partial z'} \frac{dz'}{\psi(z')R + 2x}\right).
\end{equation}

\noindent In the integrand of Eq.\,\ref{eq:totalcsrexitwakeradiation}, $\psi(z)$ is defined implicitly by the relation:

\begin{equation}
z - z' = f(\psi) = \frac{R \psi^3}{24}\frac{R \psi + 4x}{R \psi + x}.
\end{equation}

\noindent and $\Delta z_{max} = f(\phi_m)$. Here, it has been taken into account that source points positioned after the exit of the magnet do not contribute to the CSR radiation, and the first two terms arise because Eq.\,\ref{eq:totalcsrexitwakeradiation} is the result of an integration by parts. We can now follow a similar procedure to calculate the velocity component of the field given by Eq.\,\ref{eq:totalcsrfieldvelocityexit}. The kernel is strongly peaked around $\psi = 0$, and so we can assume that $\lambda(z)$ is constant over the relevant range, and apply a small-angle Taylor expansion, resulting in:

\begin{equation} \label{eq:totalcsrexitwakevelocity}
E_{||, vel}^{exit}(z, x) \approx \frac{Ne \lambda(z)}{4 \pi \epsilon_0 \gamma^2} \int_{0}^{\phi_m} \frac{2 \gamma^2}{\left(x + R\psi\right)^2}d\psi = \frac{Ne}{4\pi \epsilon_0}\frac{2\lambda(z)}{x}.
\end{equation}

\noindent This term cancels with one of the boundary terms in Eq.\,\ref{eq:totalcsrexitwakeradiation}, resulting in the following expression for the total CSR exit transient field:

\begin{equation} \label{eq:totalcsrexitwake}
E_{||}^{exit}(z, x) = E_{||, vel}^{exit}(z, x) + E_{||, rad}^{exit}(z, x) \approx \frac{Ne}{\pi \epsilon_0}\left(\frac{\lambda (z - z_{max})}{\phi_m R + 2x} + \right.
\left. \int_{z - \Delta z_{max}}^{z - \Delta z_{min}} \frac{\partial \lambda (z')}{\partial z'} \frac{dz'}{\psi(z')R + 2x}\right).
\end{equation}

\begin{figure}
\centering
\subfloat[The field lines due to the velocity field of the emitting particle ($A'$) are behind the observing particle ($A$) during rectilinear motion]{  
	\includegraphics[width=7cm]{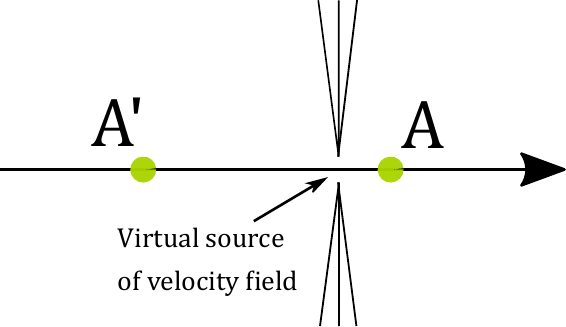} 
	\label{fig:exit_case_a}}
\hfill
\subfloat[On an arc trajectory, the velocity field lines of the emitter are in front of the observing particle.]{
	\includegraphics[width=7cm]{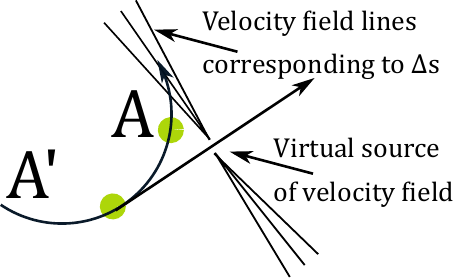}
	\label{fig:exit_case_b}}

\subfloat[In the exit transition regime, there is a small range over which the observing particle experiences a sharp spike in the CSR force due to the emitter.]{
	\includegraphics[width=8cm]{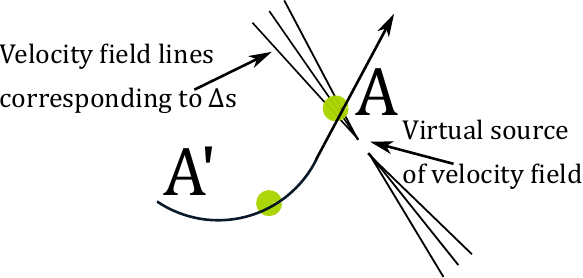}
	\label{fig:exit_case_c}}

\caption{Schematic representation of the velocity component of the CSR field on three different trajectories.}
\label{fig:exit_transient_schematic}
\end{figure}

This is equivalent to the expression for the exit transient field given in \cite{SLAC-PUB-9242}. However, we have provided a full explanation of how both the velocity and radiation components of the Li\'enard-Wiechert fields complement each other to produce this result. A comparison between Eqs.\,\ref{eq:totalcsrfieldexit}, \ref{eq:totalcsrexitwakeradiation} and the full formula \ref{eq:totalcsrexitwake} is given for a benchmark case in the following Section. For a schematic representation of the CSR velocity field during rectilinear motion, motion on an arc, and the transition regime upon exiting a curved trajectory, see Fig.\,\ref{fig:exit_transient_schematic}. A physical description for the underlying mechanism behind the interaction of both the velocity and radiation fields can be understood as follows. The contribution from the velocity field is significant only within a very small range $\psi \lessapprox \gamma^{-1} \ll 1$. The field lines corresponding to the velocity field of a relativistic particle are confined in a very flat pancake perpendicular to the direction of motion. An important property of the velocity field is that the field lines point away from a virtual source point that moves with velocity $\beta c$ in the direction that the emitter had at the time of emission. In the figure the retarded position of the emitter is shown, and the apparent, instantaneous source of the velocity field is also indicated. For a bunch moving in rectilinear motion (Fig.\,\ref{fig:exit_case_a}), this apparent source point remains coincident with the instantaneous position of the emitter. Two particles that are longitudinally next each other barely feel each other's field due to the pancake effect. The field lines of the upstream particle are always behind the downstream particle. 

In case of an arc (Fig.\,\ref{fig:exit_case_b}), the path of the observer curves away from the direction that the emitter had at time of emission (denoted the \lq$z$ direction\rq). Therefore the $z$ component of the velocity of the observer becomes lower than $\beta c$ during the transit time in which the field travels from emitter to observer. Therefore, at time of observation, the \lq pancake region\rq\,of dense field lines is in front of the observer. In either the rectilinear or arc case, the upstream particle observes a very small field. However, at the end of the arc (Fig.\,\ref{fig:exit_case_c}), the geometry must pass from a situation with field lines in front of the observer to a situation with field lines behind the observer. Hence there must be a point where this field passes over the observer, giving a spike of CSR force. This effect is the exact analogue of the entrance effect sketched in Fig.\,\ref{fig:entrance_transient_schematic}, in which the geometry transits from a case with the velocity field behind the observer to a situation with the field in front of it.

The above mechanism may also explain why the contribution of the velocity field is only significant in the very final angular range $0 < \psi < \gamma^{-1}$ of the arc, as detailed in the previous section. This is simply the angular extent of the pancake field that needs to pass over the observing particle. It should also be noted that the field line patterns sketched in the figures are not entirely realistic, because there will only be a thin radiation shell of thickness $\Delta s/\beta$ generated from the path element $\Delta s$, and only in this thin shell the drawn pancake field line pattern exists. The subsequent path element will generate another radiation shell, and the corresponding \lq\,pancake field\rq\,inside that shell will be slightly differently oriented due to the different orientation of the path element. The total field line pattern will be the sum of all such infinitesimal shells-with-pancake-fields, in which the concept of a pancake field will be hard to recognize at all. The main point is, however, that with any path element there is an associated region of dense field lines. Near the end of the arc, there is a point where this region will pass over the upstream particles, creating a brief but intense spike of CSR force.

\section{Numerical Validation}\label{sec:numericalvalidation}

\begin{figure}
	\begin{center}
		\centering           
		\includegraphics{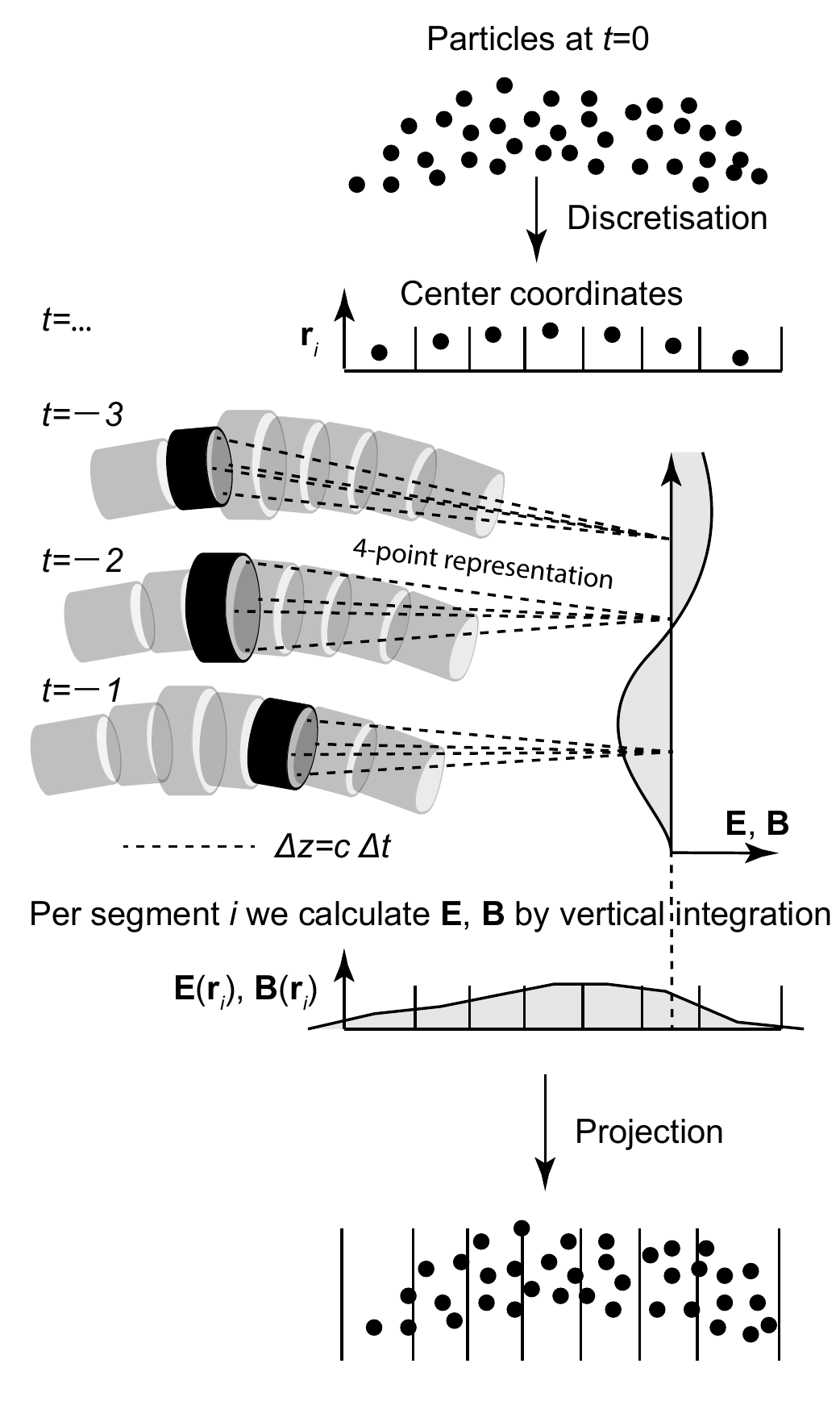}
		\caption{Representation of particle bunch adopted by \textsc{GPT} to calculate CSR forces. The particle bunch is discretised and represented by slices. The CSR force is then calculated and projected onto the bunch.} 
		\label{fig:gptcsr}
	\end{center}
\end{figure}

In order to validate the analytical results of Sec.\,\ref{sec:entrancetransient} and \ref{sec:exittransient}, we have numerically calculated the electromagnetic field distribution in an electron bunch in both the entrance and exit transient regimes using the \textsc{GPT} code \cite{GPT}. \textsc{GPT} is a particle tracking code that integrates the equations of motion of a large number of charged particles in the presence of electromagnetic fields. The code has the option to include the computation of the retarded Li\'enard-Wiechert fields of the tracked particles. Because this involves the storage of the trajectory of the particles and solution of retardation conditions, calculation of Li\'enard-Wiechert fields is computationally expensive. To reduce the computational cost, the \textsc{GPT} code does not evaluate the field of each tracked particle, but instead represents the particle bunch by a number of bunch slices (see Fig.\,\ref{fig:gptcsr}). Each bunch slice is represented by either four or sixteen off-axis particles that are spaced according to the RMS transverse size of the slice in order to capture the impact of the transverse extent of the bunch. While integrating the equation of motion of a tracked particle, \textsc{GPT} evaluates the Li\'enard-Wiechert field resulting from the past trajectory of each of the representative particles at the longitudinal position of the tracked particle. It is important to note that \textsc{GPT} uses the exact expression for the Li\'enard-Wiechert fields based on the numerically obtained coordinates of particles in the bunch, and does not apply any analytic approximation or presumed trajectory of the bunch. The parameters used in the simulation are listed in Table\,\ref{table:gptsimulationparameters}. We deliberately chose artificially small energy spread and transverse bunch size, and used hard-edged magnet fringes in the exit transient simulations to match the analytic case as much as possible.

\begin{table}
\caption{\label{table:gptsimulationparameters}Lattice and electron bunch parameters used in the \textsc{GPT} simulation -- used for validation of the numerical study.}
\begin{centering}
\begin{tabular}{|c|c|c|}
	\hline
	\textbf{Lattice} & \textbf{Value} & \textbf{Unit} \\ [0.5ex] 
	\hline 
	Magnet length $R\phi_m$    & 1.14 & \si{\metre} 			   \\
	Radius of curvature $R$    & 2.29 & \si{\metre} 			  \\ 
	Drift length before bend   & 0.1, 50 & \si{\metre} 			   \\
	Entrance / exit edge angle & 0 & \si{\radian} 			   \\
	Fringe width (entrance)    & 1.7 & \si{\milli\metre} 			   \\ 
	Fringe width (exit)		   & 0 & \si{\milli\metre} 			   \\ 
	\hline\hline
	\textbf{Initial bunch}     &  						  & \\
	\hline
	Number of macroparticles   & $10^6$ &  						   \\
	Bunch charge 			   & 70 & \si{\femto\coulomb} 	   \\
	Mean energy 			   & 380 & \si{\mega\electronvolt}   \\
	Twiss $\beta_x$ 		   & 1.34  & \si{\metre} 			   \\
	Twiss $\beta_y$ 		   & 3 & \si{\metre} 			   \\
	Twiss $\alpha_x$ 		   & 0.185 & \si{\radian} 			   \\
	Twiss $\alpha_y$ 		   & 0 & \si{\radian} 			   \\
	$\epsilon_{N,x,y}$         & $5 \times 10^{-3}$ & \si{\micro\metre\radian}  \\
	RMS bunch length 	       & 0.9 & \si{\metre} 			   \\
	Uncorrelated energy spread & 0 & 						  \\
	Energy chirp ($dE/dz$)	   & 0 & $\%/\si{\milli\metre}$    \\
	\hline
\end{tabular}
\end{centering}
\end{table}

\subsection{Entrance Transient Effect}

\begin{figure}
\begin{center}
	\centering           
	\includegraphics[width=7.7cm]{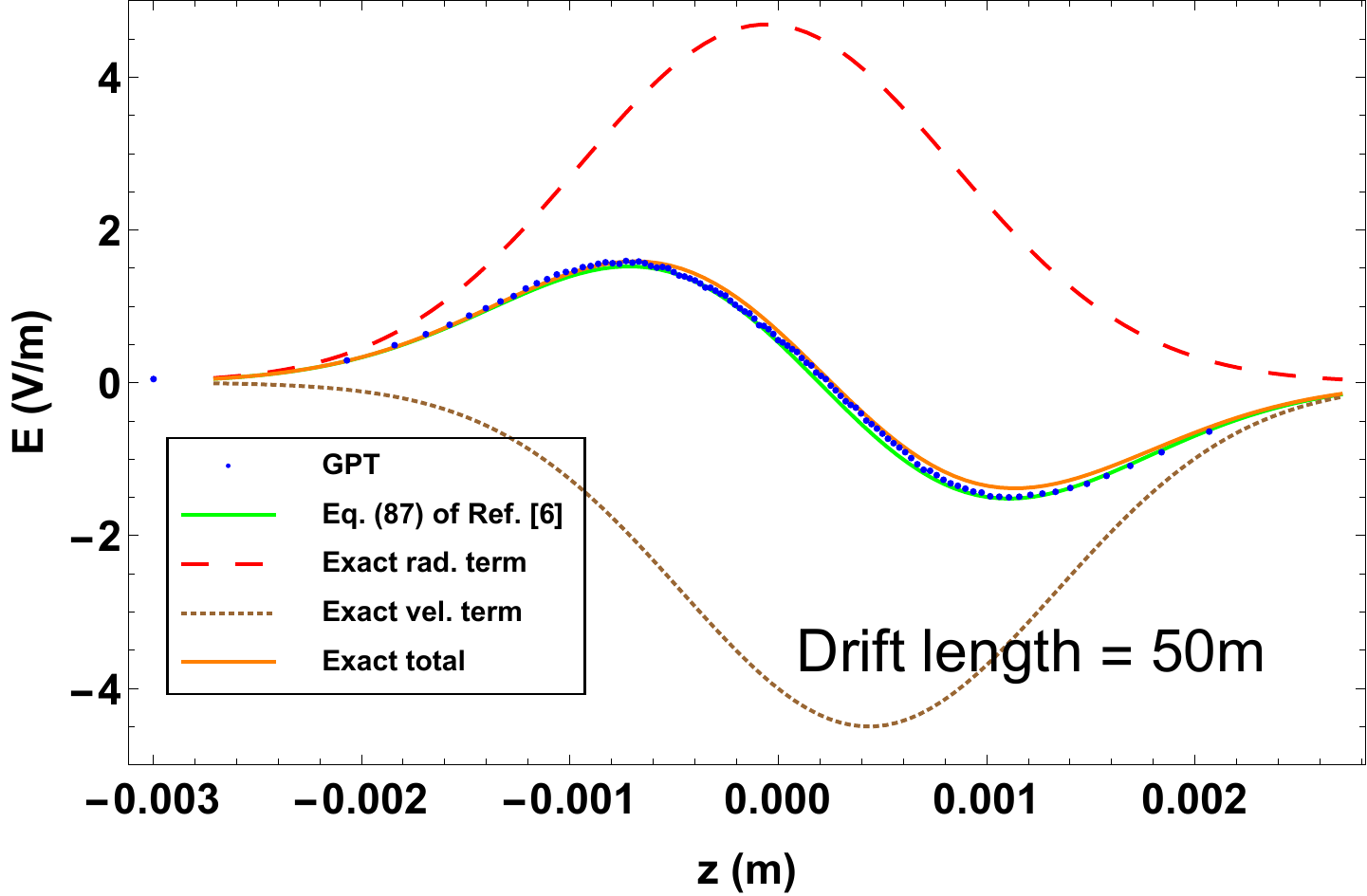}
	\hfill
	\includegraphics[width=7.7cm]{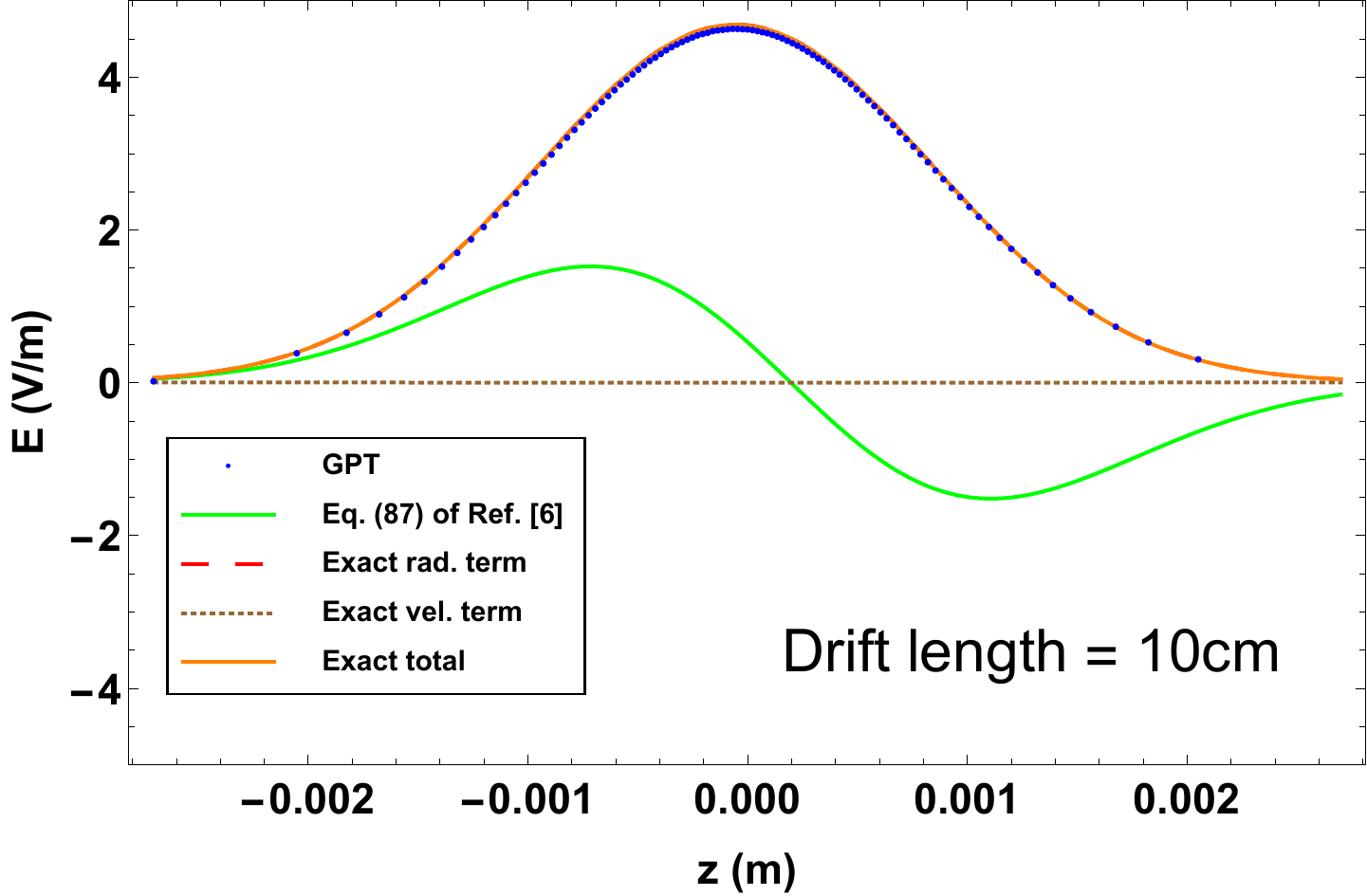}
	\caption{Longitudinal component of CSR electric field as a function of longitudinal position in the bunch for the parameters in Table\,\ref{table:gptsimulationparameters} and a drift before the magnet of: Left: $50$\,\si{\metre}; and Right: $10$\,\si{\centi\metre}, as simulated by \textsc{GPT}, against both Eq.\,[87] of Ref.\,\cite{NIMA.1997.2.373} and Eq.\,\ref{eq:entrancetransientfield} -- both the velocity and radiation terms individually, and combined. Positive values of $z$ refer to the head of the bunch.}
	\label{fig:entrance_transient}
\end{center}
\end{figure}

The CSR field was initially calculated by \textsc{GPT} at a point $24$\,\si{\centi\metre} into the magnet in order to simulate the entrance transient field. This distance is only half that of the steady-state condition $D^{SS}$ (Eq.\,\ref{eq:steadystatecondition}), and so it is expected that the general expression of Eq.\,\ref{eq:steadystatefield} will be required to calculate the fields. In this simulation, the drift before the magnet was set to $50$\,\si{\metre}. The results from the simulation are in good agreement with Eq.\,\ref{eq:steadystatefield}, as seen in the left-hand plot of Fig.\,\ref{fig:entrance_transient}. However, if the simulation is run again, but with the drift before the bend set to $10$\,\si{\centi\metre}, the \textsc{GPT} result effectively reduces to Eq.\,\ref{eq:steadystatefield}, and thereby differs from the usual approximation of Ref.\,\cite{NIMA.1997.2.373}. The second term on the right-hand side of Eq.\,\ref{eq:entrancetransientfield} is suppressed by lowering the integration boundary, showing that the approximation of an infinitely long drift before the entrance to a bending magnet may not be valid for some cases. As shown in the right-hand plot of Fig.\,\ref{fig:entrance_transient}, the \textsc{GPT} simulation reflects this behaviour.

\subsection{Exit Transient Effect}\label{sec:transitionregime}

Having found a qualitatively different behaviour of the electric field inside and past the bending magnet, it is of interest to study the transition from one regime to the other. Equation \ref{eq:totalcsrfieldexit} assumes that all particles follow the same reference trajectory. However, the impact of a transverse displacement of the emitting electrons on the observed electric field may be studied by including a vertical offset (out of paper) of the emitting electron in  Fig.\,\ref{fig:csr_exit_transient}. Figure \ref{fig:sideview} gives a side view of the resulting configuration. Due to the offset, the distance $\sigma$ from emitter to observer becomes:

\begin{figure}
	\begin{center}
		\centering           
		\includegraphics[width=8cm]{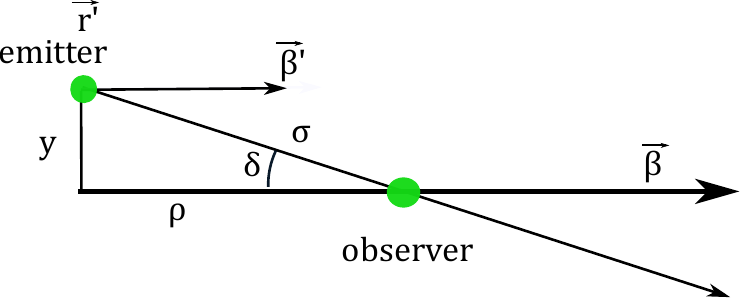}
		\caption{Side view of the configuration of Fig.\,\ref{fig:csr_exit_transient} in case of a vertical offset $y$ of the emitting particle.}
		\label{fig:sideview}
	\end{center}
\end{figure}

\begin{equation}\label{eq:ysubstitution}
\sigma = \sqrt{\rho^2 + y^2} = \sqrt{4R^2 \sin{\psi/2}^2 + 2Rx\sin{\psi} + x^2 + y^2}
\end{equation}

\noindent In addition, the angles $\theta$, $\eta$ and $\xi$ are stretched somewhat, such that their cosines become smaller by a factor $\cos{\delta} = \rho/\sigma$. Re-evaluating Eq.\,\ref{eq:lwexpandedfield} with these modifications shows that the electric field is still given by Eq.\,\ref{eq:totalcsrfieldexit} after the substitution $\rho\to\sigma$.

\begin{figure}
	\begin{center}
		\centering           
		\includegraphics[width=15cm]{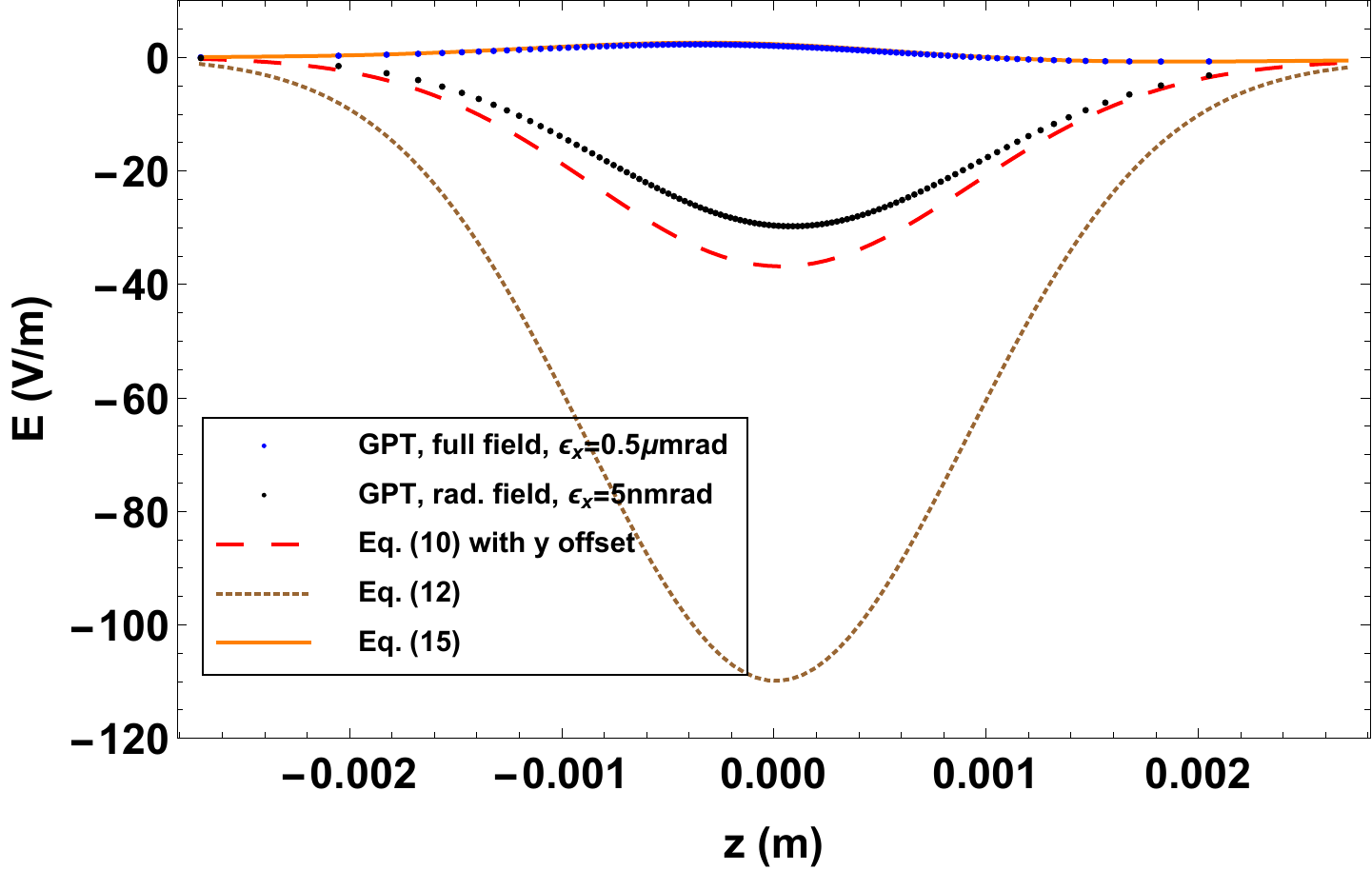}
		\caption{Longitudinal component of electric field as a function of longitudinal position in the bunch, at $5$\,\si{\milli\metre} from the exit of the dipole. Blue: \textsc{GPT} simulation of the full field; Black: \textsc{GPT} simulation of the radiation field only; Red: Eq.\,\ref{eq:totalcsrfieldexit} with an offset in the $y$ plane; Brown: Eq.\,\ref{eq:totalcsrexitwakeradiation}; Orange: Eq.\,\ref{eq:totalcsrexitwake}.}
		\label{fig:longitudinalefield}
	\end{center}
\end{figure}

Fig.\,\ref{fig:longitudinalefield} shows the longitudinal component of the electric field as a function of longitudinal position in the bunch evaluated at $5$\,\si{\milli\metre} past the bending magnet. The results for the GPT simulation of both the full CSR field, and the radiation component, are compared with Eqs.\,\ref{eq:totalcsrexitwakeradiation} and \ref{eq:totalcsrexitwake}, and Eq.\,\ref{eq:totalcsrfieldexit} with an offset in the $y$ plane according to Eq.\,\ref{eq:ysubstitution}. The simulation results agree well with the expression  Eq.\,\ref{eq:totalcsrfieldexit}, with the inclusion of a small transverse offset. The fact that the approximation for the radiation term Eq.\,\ref{eq:totalcsrexitwakeradiation} differs greatly from both the exact formula for the radiation field and the simulation results demonstrate the importance of including the velocity term when computing CSR fields at the exit of a bending magnet. Fig.\,\ref{fig:longitudinalefield} shows that Eq.\,\ref{eq:totalcsrexitwake} fully captures the actual behaviour of the field that we found both analytically and numerically. 

\section{Parameter Scans} \label{sec:parameterscans}

\begin{figure}
	\begin{center}
		\centering           
		\includegraphics[width=14cm]{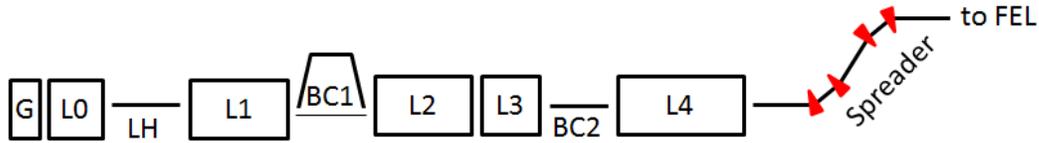}
		\caption{Sketch, not to scale, of the FERMI linac to FEL beam line. This study applies from the Gun (G) through the first accelerating sections (L0 and L1) and the laser heater (LH) to the exit of the first bunch compressor (BC1).}
		\label{fig:fermi_linac}
	\end{center}
\end{figure}

A schematic of the FERMI linac is shown in Fig.\,\ref{fig:fermi_linac}. The emittance was measured at the exit of the first bunch compressor, BC1, as a function of Linac 1 RF phase (i.e. energy chirp, that is, a longitudinal energy-to-position correlation along the bunch), chicane bending angle, and the strength of the last quadrupole before the entrance to the bunch compressor. The first two scans implied a scan of the bunch length compression factor in the range $20 - 64$ and $8 - 60$ for the Linac 1 phase and chicane bending angle scans, respectively. The scan of quadrupole strength was done at the fixed compression factor of $36$. The compression process was kept linear during the scan by virtue of an X-band RF cavity, which allows to approximately preserve the current shape through the chicane, as shown later in Figs.\,\ref{fig:peak_current_bc01} and \ref{fig:peak_current_l01} \cite{LCLS-TN-01-1,IPAC2010.TUPE015}. During the phase scan, the accelerating gradient of Linac 1 was scaled in order to keep the mean bunch energy constant at the entrance to BC1. Measurements were taken using the single quad-scan technique \cite{SLAC-R-621}, by varying the strength of one quadrupole magnet (Q\_BC01.07), located in the section directly after BC1. The machine was operated with a constant bunch charge of $100$\,\si{\pico\coulomb}, and a mean energy of approximately $300$\,\si{\mega\electronvolt} at BC1. For each set of scans, the two other scanning parameters were kept constant. During the experimental run, the following scans were performed:

\begin{itemize}
\item Linac 1 phase -- vary between $70.5$\,\si{\degree} and $73.3$\,\si{\degree} (nominal is $73$\,\si{\degree}).
\item BC1 angle -- vary between $100$\,\si{\milli\radian} and $109$\,\si{\milli\radian} (nominal is $105$\,\si{\milli\radian}).
\item Q\_L01.04 K1 (this is the final quadrupole before the entrance to BC1) -- vary between $-2.0$\,\si{\per\metre\squared} and $5.0$\,\si{\per\metre\squared} (nominal is $1.6$\,\si{\per\metre\squared}).
\end{itemize}

\begin{table}
\caption{\label{table:fermi_parameters} Main beam parameters of the FERMI accelerator at the entrance to BC1 in the nominal configuration.}
\begin{centering}
\begin{tabular}{|c|c|c|}
	\hline
	\textbf{Bunch parameters}  & \textbf{Value} & \textbf{Unit}\\
	\hline
	Bunch charge 			   & $100$ & \si{\pico\coulomb} 	   \\
	Mean energy 			   & $300$ & \si{\mega\electronvolt}   \\
	$\epsilon_{N,x,y}$         & $0.62$ & \si{\micro\metre\radian}  \\
	RMS bunch length 	       & $0.61$ & \si{\milli\metre} \\
	Relative energy spread	   & $0.95$ & $\%$ \\
	Distance between 1\textsuperscript{st} - 2\textsuperscript{nd}, and 3\textsuperscript{rd} - 4\textsuperscript{th} bend	   & $2.5$ & \si{\metre} \\
	Distance between 2\textsuperscript{nd} and 3\textsuperscript{rd} bend	   & $1.0$ & \si{\metre} \\
	Momentum compaction $R_{56}$ & $0.057$ & \si{\metre} \\
	s-E correlation $\frac{1}{E_0}\frac{dE}{ds}$			   & $-17.0$ & \si{\metre}\textsuperscript{-1} \\
	\hline
\end{tabular}
\end{centering}
\end{table}

At the diagnostic stations, both Yttrium Aluminum Garnet (YAG) and Optical Transition Radiation (OTR) screens are available. Estimates of the resolution for these screens are, respectively, $45$\,\si{\micro\metre} for a pixel width of $31.2$\,\si{\micro\metre} and $25$\,\si{\micro\metre} for a pixel width of $19.6$\,\si{\micro\metre}\,\cite{JINST.8.P05015}. The majority of the measurements were initially taken with OTR screens, but coherent effects were suspected to be having an effect on the measured emittance as the bunch approached maximum compression, and so some measurements were repeated with YAG screens, whose performance is expected to suffer much less from coherent emission. The typical emittance measurement procedure consists of taking at least 5 images for between $10$ and $30$ settings of Q\_BC01.07, and the single quad-scan technique (see, for example, \cite{SLAC-R-621}) is applied to calculate the transverse emittances and Twiss parameters. 

\section{Simulation Setup} \label{sec:simulationsetup}

\begin{figure}[h]
	\begin{center}
		\centering           
		\includegraphics[width=8cm]{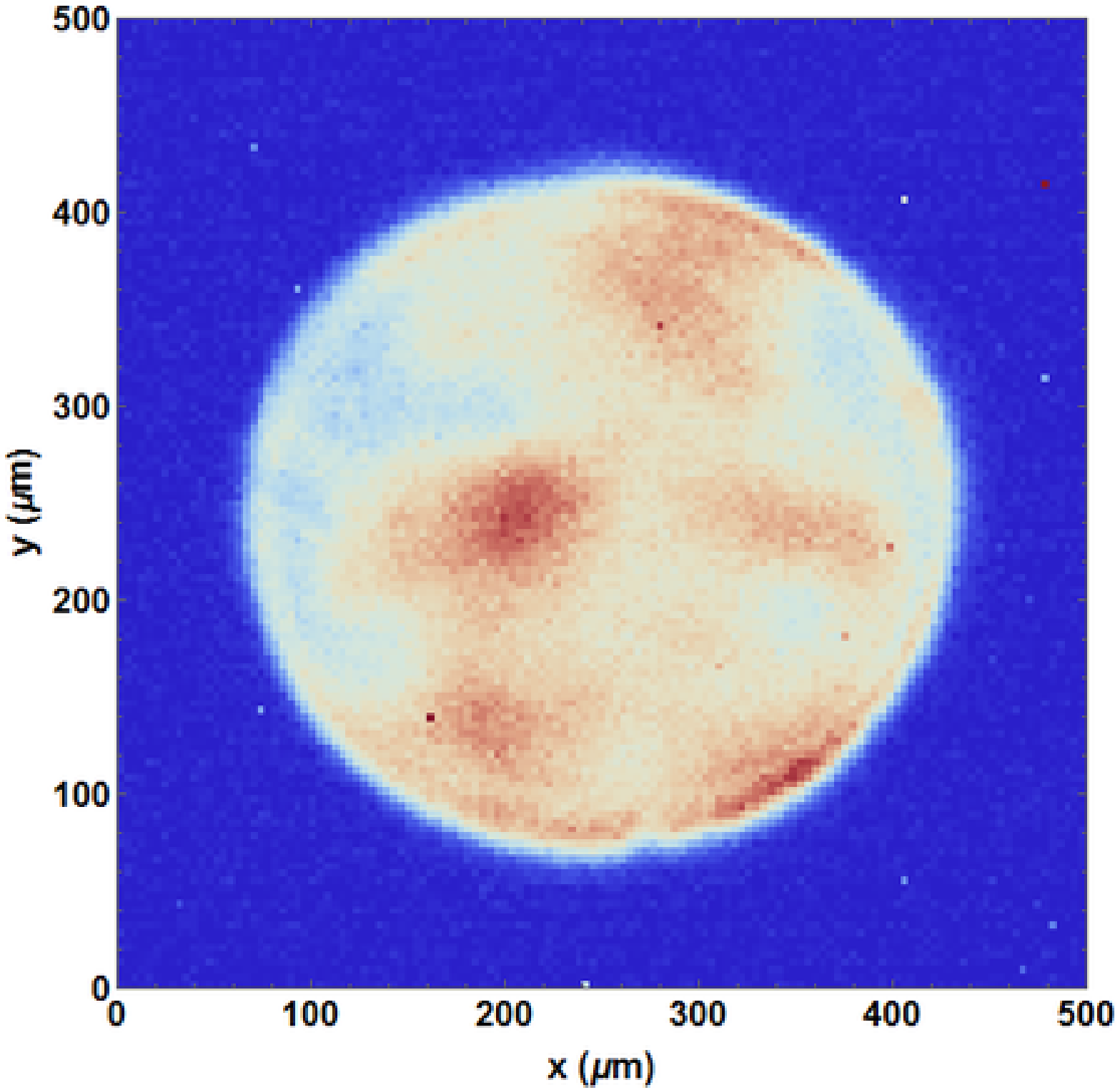}
		\caption{Measured transverse photo-injector laser profile used as input for \textsc{GPT} simulation.}
		\label{fig:transverse_laser_spot}
		\centering           
		\includegraphics[width=8cm]{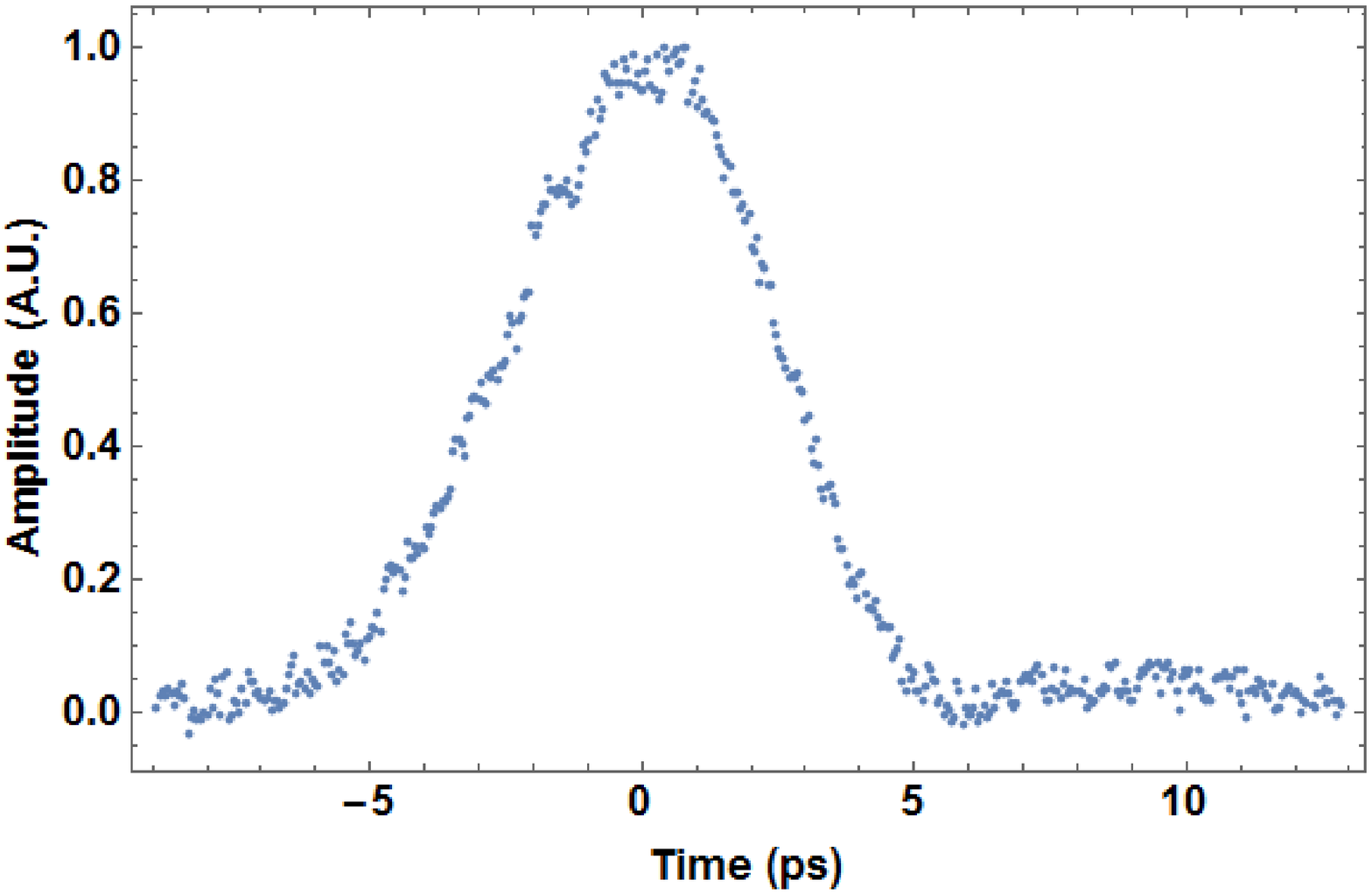}
		\caption{Measured longitudinal photo-injector laser profile used as input for \textsc{GPT} simulation.}
		\label{fig:longitudinal_laser_spot}
	\end{center}
\end{figure}

Simulations of the FERMI injector (up to the exit of the first linac, see Fig.\,\ref{fig:fermi_linac}) have been produced using \textsc{GPT}. In order to accurately match the simulation to experimental conditions, the measured transverse and longitudinal profiles of the photoinjector laser were used as input parameters to the simulation (shown in Figs.\,\ref{fig:transverse_laser_spot} and \ref{fig:longitudinal_laser_spot}), along with geometric wakefields from the injector linac. Full 3D space-charge effects were also included. The injector linac phase was optimised for minimal energy spread -- as is done in the routine procedure of linac tuning -- and good agreement was found between the simulated and experimentally measured bunch properties at the exit of the injector.

From this injector simulation, the bunch was then tracked using the \textsc{Elegant} code \cite{APS-LS-287} up to the entrance of BC1, including the effects of linac wakefields, the laser heater, which is a tool aimed to suppress the so-called microbunching instability that otherwise develops as the bunch propagates through the accelerator \cite{NIMA.2004.1.355,PhysRevSTAB.17.120705}, coherent synchrotron radiation and space charge models. From this point, three particle tracking codes have been used to compare the emittance measurement results with simulation: \textsc{Elegant}, \textsc{CSRTrack} \cite{FEL2004.MOCOS05} and a modified version of \textsc{GPT} which utilises the CSR model outlined above in Sec.\,\ref{sec:numericalvalidation}. In the 1D CSR simulations, \textsc{Elegant} applies the calculation of Saldin \textit{et al} \cite{NIMA.1997.2.373} to calculate the energy change due to coherent radiation in a bend, and the subsequent transient effect some time after the bunch exits the dipole, based on \cite{SLAC-PUB-9242}. In the \textsc{Elegant} calculation, the dipole is split up into pieces, and the bunch is tracked sequentially through each piece. At each point, the bunch is projected onto the reference trajectory and the electric field of the bunch is computed. The projected (1D) method of \textsc{CSRTrack} divides the bunch up into Gaussian \lq sub-bunches\rq, smooths the distribution, and calculates the CSR field from a convolution of the distribution with a kernel function describing Li\'enard-Wiechert fields across the bunch trajectory \cite{TESLA-FEL-Report-2003-05}. At the exit of the bunch compressor (including a drift to account for transient CSR effects), the \textsc{CSRTrack} output is then converted back into \textsc{Elegant}, and tracked up to Q\_BC01.07, the measurement point.

As described above in Sec.\,\ref{sec:numericalvalidation}, the CSR routine in \textsc{GPT} does not employ the 1D approximation, but calculates the retarded Li\'enard-Wiechert potentials directly by slicing up the bunch longitudinally, and it does not directly project the radiating particles onto a line. Each slice contains a number of radiation emission points (typically four), and the full history of both the fields and the particle coordinates are stored for each time step. The 3D routines in \textsc{CSRTrack} also calculate the radiation fields directly, but in a slightly different manner. For our simulations, we have utilised the \code{csr\_g\_to\_p} method, in which the particles are first replaced by Gaussian \lq sub-bunches\rq, and the radiation field is calculated via a pseudo-Green's function approach \cite{TESLA-FEL-Report-2003-05}, with each sub-bunch having an effect on each particle in front of it.

It should be mentioned that the number of bins used for the density histogram in the CSR and longitudinal space-charge (LSC) models of \textsc{Elegant}, in addition to the smoothing applied on the bunch, can have an impact on the final results \cite{PhysRevSTAB.4.070701}. Following a convergence study, by varying the number of CSR bins between $100$ and $5000$, and performing the parameter scans for $10^6$, $5 \times 10^6$ and $10^7$ macroparticles, we have determined that $500$\,LSC and CSR bins for an \textsc{Elegant} simulation of $10^6$ macroparticles is sufficient. Following previous studies \cite{FERMILAB-TM-2533}, we set the sub-bunch size for the \textsc{CSRTrack} calculations to be $10$\,\si{\percent} of the rms bunch length. A similar set of simulations was run in \textsc{GPT} in order to achieve convergent results, for input distributions of $10^5$, $10^6$ and $10^7$ macroparticles. To determine the significance of the Coulomb term outlined in Sections \ref{sec:entrancetransient} and \ref{sec:exittransient}, \textsc{GPT} simulations were also run with this term deactivated. Since dipole fringe fields are included in \textsc{CSRTrack} by default, the parameter scans were also simulated with dipole fringes in the other two codes. This should also provide the most realistic benchmark with the experimental case. 

\section{Results} \label{sec:results}

\begin{figure}[h]
	\begin{center}
	\subfloat[Results from the BC01 angle scan.]{  
		\includegraphics[width=7.7cm]{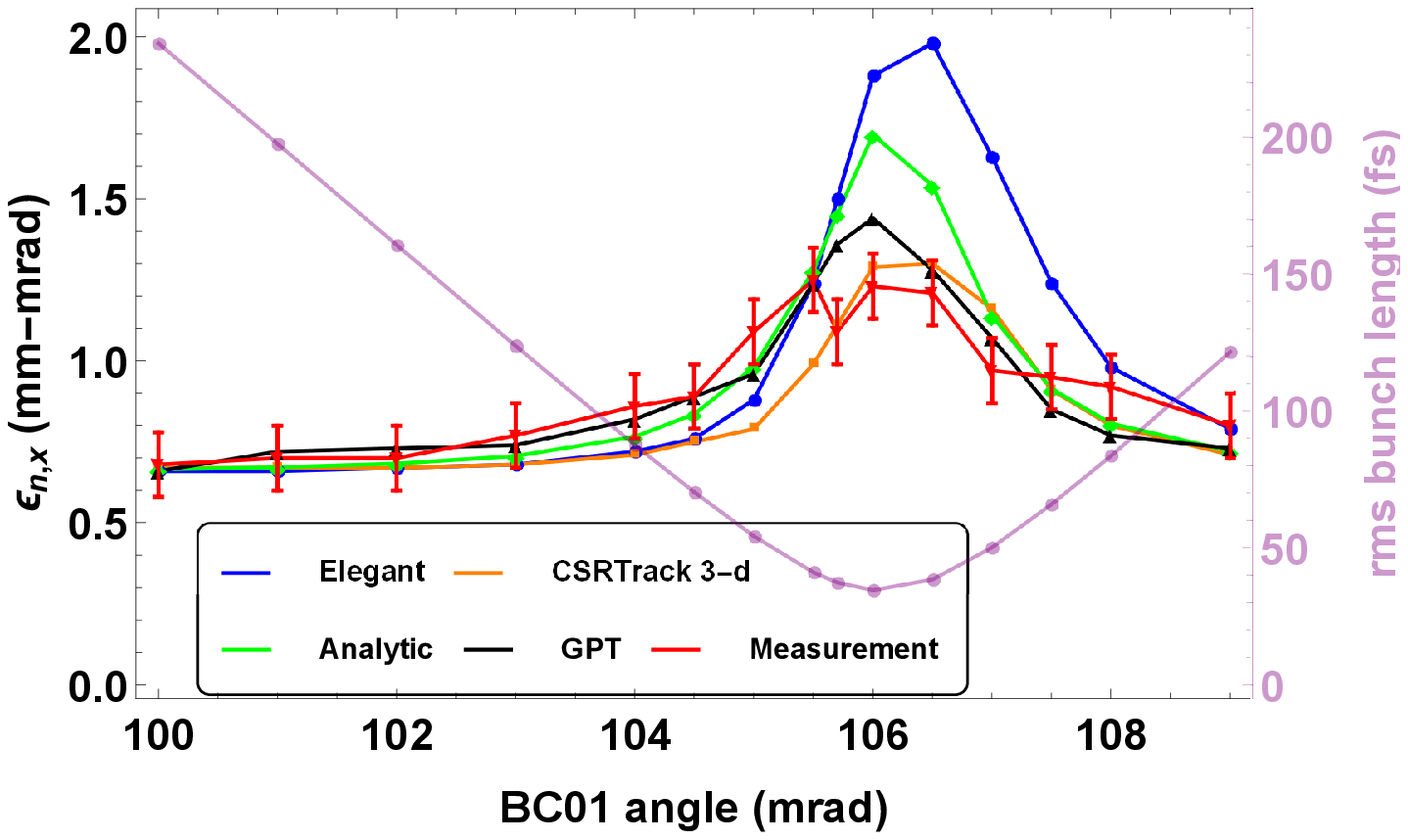} 
		\label{fig:emit_bc01}}
	\hfill
	\subfloat[$D_{par}$ at the exit of each bunch compressor dipole for the linac phase scan.]{
		\includegraphics[width=7cm]{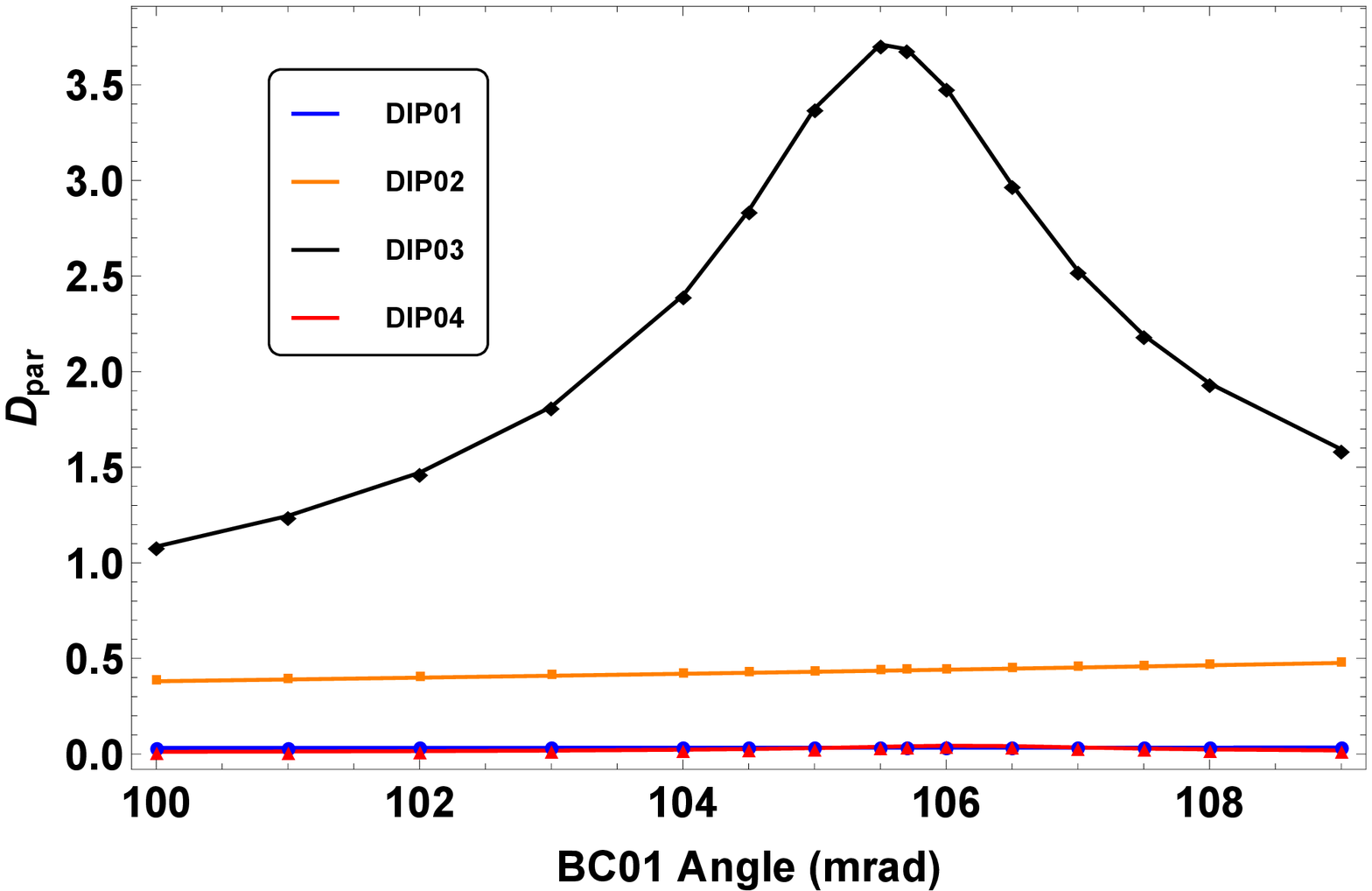}
		\label{fig:derbenev_bc01}}
	\caption{Horizontal emittance as a function of BC01 bending angle, with the corresponding bunch length as simulated by \textsc{Elegant}. The analytic results are calculated using Eq.\,\ref{eq:emit_growth_case_b}.}
	\end{center}
\end{figure}

\begin{figure}
	\begin{center}
		\subfloat[Results from the L01 phase scan.]{  
			\includegraphics[width=7.9cm]{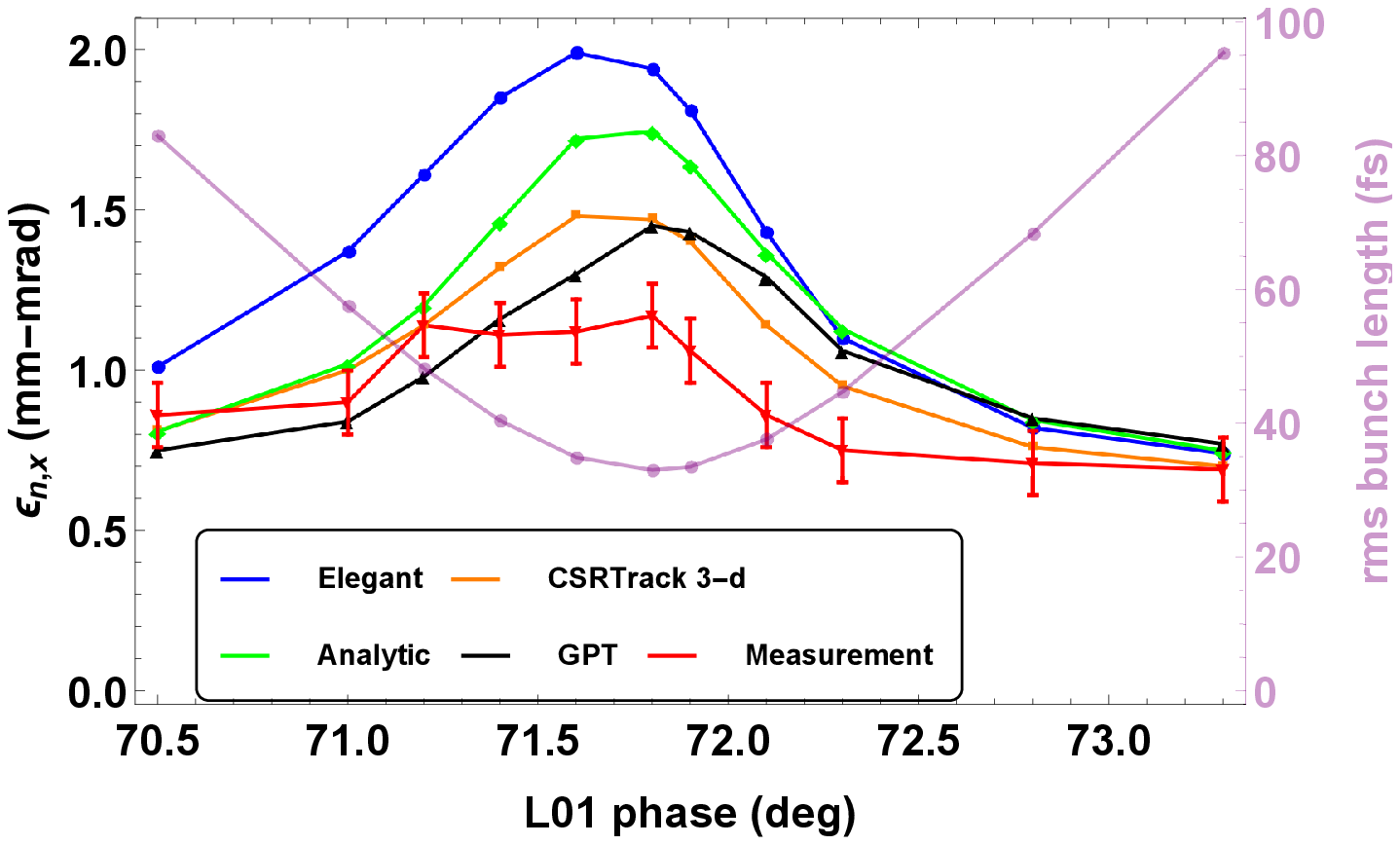} 
			\label{fig:emit_l01}}
		\hfill
		\subfloat[$D_{par}$ at the exit of each bunch compressor dipole for the linac phase scan.]{
			\includegraphics[width=7cm]{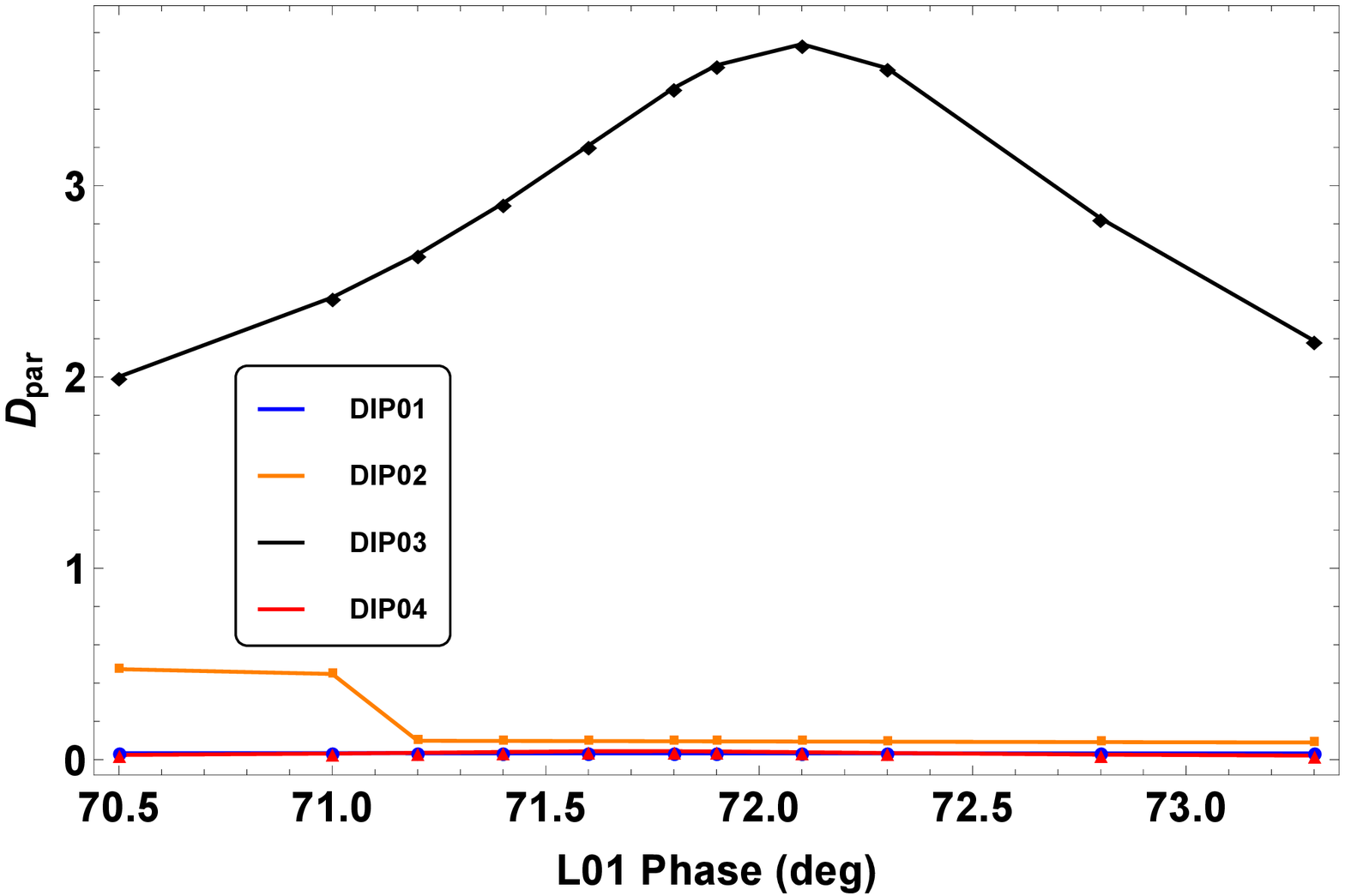}
			\label{fig:derbenev_l01}}
		\caption{Horizontal emittance as a function of Linac 1 phase, with the corresponding bunch length as simulated by \textsc{Elegant}. The analytic results are calculated using Eq.\,\ref{eq:emit_growth_case_b}.}
	\end{center}
\end{figure}

\begin{figure}
	\begin{center}
		\subfloat[Results from the Q\_L01.04 scan.]{  
			\includegraphics[width=7.5cm]{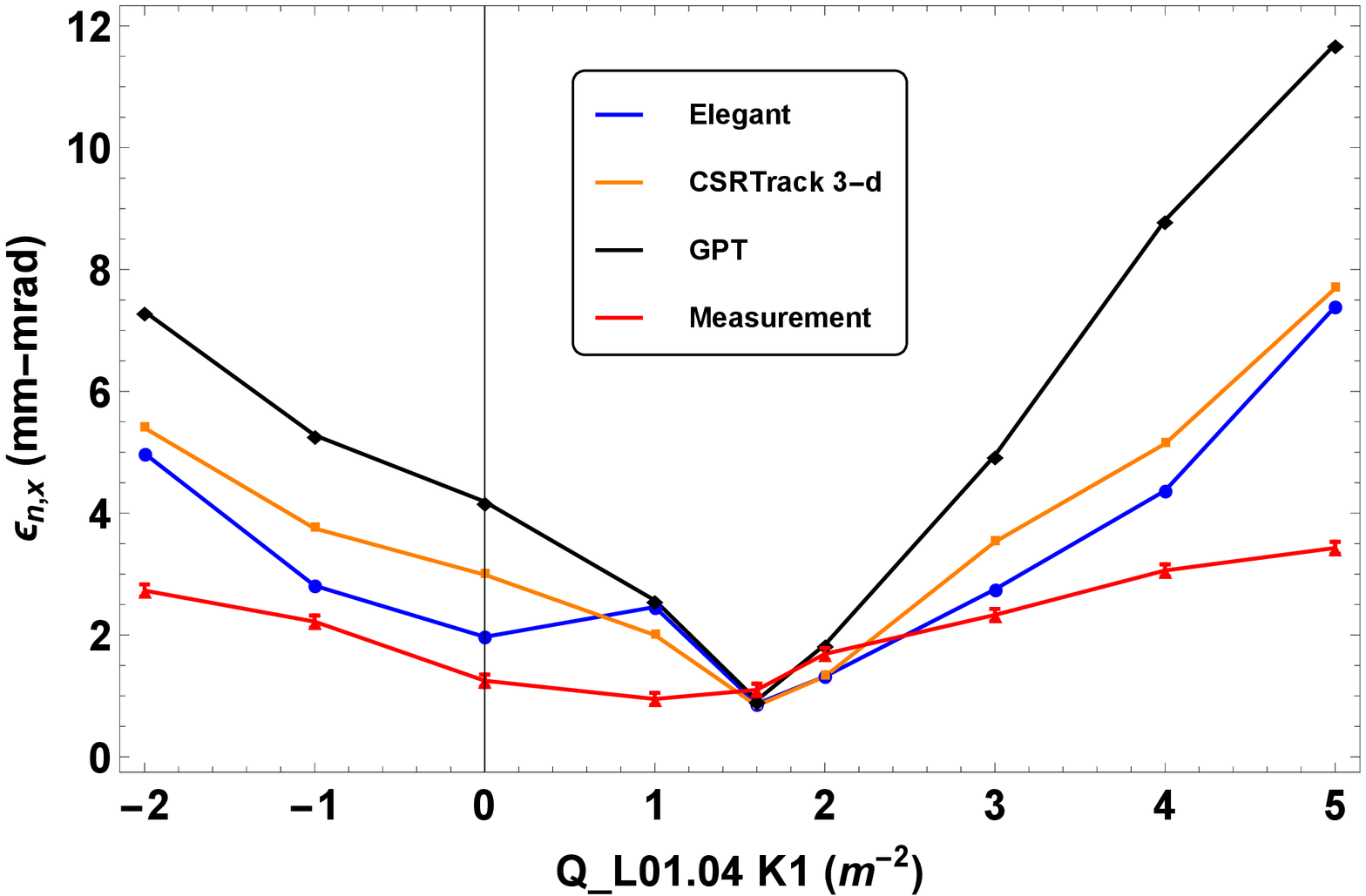} 
			\label{fig:emit_quadscan}}
		\hfill
		\subfloat[$D_{par}$ at the exit of each bunch compressor dipole for the linac phase scan.]{
			\includegraphics[width=7.5cm]{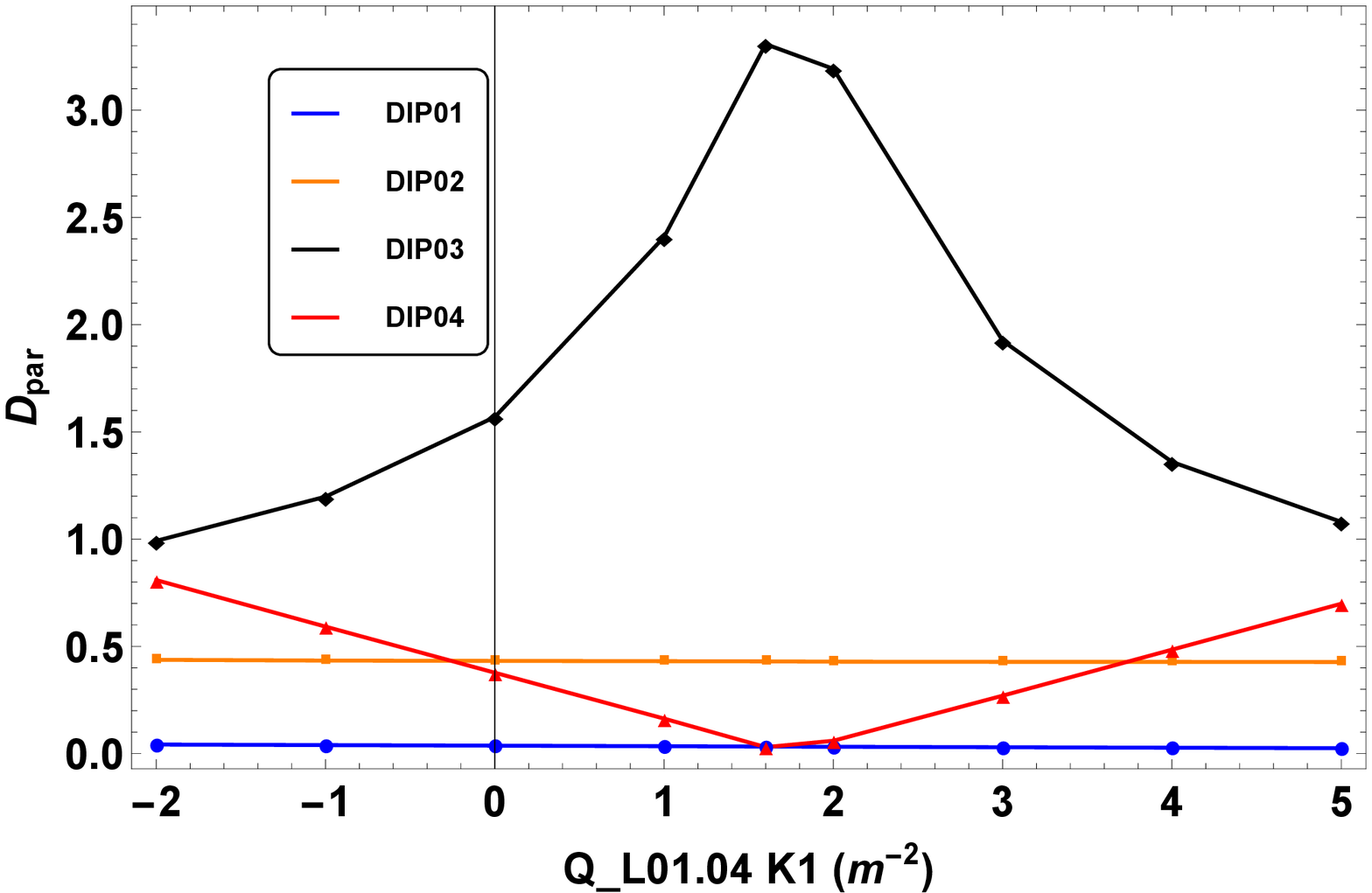}
			\label{fig:derbenev_q0104}}
		\caption{Horizontal emittance as a function of Q\_L01.04 strength}
	\end{center}
\end{figure}

During the experimental run, the parameter scans detailed in Sec.\,\ref{sec:parameterscans} were performed, and the emittance was measured by quad scan using the FERMI online emittance tool. Plots comparing the emittance measurements with simulation results from the two codes are given in Figs.\,\ref{fig:emit_bc01},\ref{fig:emit_l01} and \ref{fig:emit_quadscan}. The CSR-induced emittance growth in these regimes has also been calculated, based on the 1D analytic theory given in \cite{SLAC-PUB-8028}. The emittance growth corresponding to the longitudinal and transverse CSR wake with the entire bunch travelling on a circular orbit (i.e. the steady-state regime) are given as \cite{SLAC-PUB-8028,PhysRevAccelBeams.20.064402}:

\begin{subequations}\label{eq:emit_growth_case_b}
	\begin{gather}
	\Delta \epsilon_{N}^{long} = 7.5\times 10^{-3}\frac{\beta_x}{\gamma} \left(\frac{r_e N {L_b}^2}{R^{5/3}\sigma_z^{4/3}} \right)^2 \\
	\Delta \epsilon_{N}^{trans} = \frac{\left(-3+2\sqrt{3}\right)}{24\pi}\frac{\beta_x}{\gamma} \left(\frac{\Lambda r_e N L_b}{R\sigma_z} \right)^2,
	\end{gather}
\end{subequations}

\noindent with $\beta_x$ the horizontal beta function, and

\begin{equation}
\Lambda = \ln\left(\frac{(R \sigma_z^2)^{2/3}}{\sigma_x^2}\left(1+\frac{\sigma_x}{\sigma_z}\right)\right).
\end{equation} 

We also provide calculations of the Derbenev parameter $D_{par}$ \cite{TESLA-FEL-Report-1995-05} in Figs.\,\ref{fig:derbenev_bc01}, \ref{fig:derbenev_l01} and \ref{fig:derbenev_q0104}, in order to illustrate the that the validity range of the 1D CSR approximation is violated approaching maximal compression or in cases where the transverse beam size is large. For the analytical calculations to be valid, the condition $D_{par} << 1$ should be fulfilled. This parameter is given by:

\begin{equation}
D_{par} = \sigma_\perp \sigma_z ^{-2/3} R^{1/3}.
\end{equation}

\noindent The values for the transverse beam size $\sigma_\perp$ and bunch length are taken from \textsc{Elegant} simulations with CSR switched off.

As seen from the plots, there is a general agreement between the measurement procedure, simulation results and analytic calculations, at least in terms of the trends. For the quad scans (Fig.\,\ref{fig:emit_quadscan}) in particular, some post-processing was necessary in order to crop some of the images -- for a strongly mismatched bunch, some of the bunches were barely visible above the noise. The discrepancy between simulation and experiment in the peak around $71.6 - 72.1$\si{\degree} in Fig.\,\ref{fig:emit_l01} can be attributed to coherent OTR emission (COTR). It has been demonstrated elsewhere  \cite{PhysRevSTAB.11.030703,PhysRevSTAB.12.040704} that intense COTR emission can lead to an underestimation of the transverse beam size, and thereby to lower measured emittance values. In both sets of experimental data for varying compression factor, there is a slight dip around the point where a peak in emittance is seen in simulation. This interpretation is supported by the observation that the largest mismatch between the \textsc{Elegant} simulation and the two sets of experimental results for this data set occur where the bunch length is around $40$\,\si{\femto\second} or less -- this is the point where coherent emission is expected to be maximised. We observe a similar apparent overestimation of emittance growth for the bunch compressor angle scan in Fig.\,\ref{fig:emit_bc01} for the \textsc{Elegant} simulation. \textsc{GPT} and \textsc{CSRTrack} 3D are able to capture both the emittance trend and its absolute value more accurately over the entire range of bunch lengths. It is also possible, however, that shielding of CSR by the vacuum pipes in the bunch compressor could contribute to the mismatch between the simulated and experimental results -- this was not a factor included in any of the simulations.

\begin{figure}[h]
	\begin{center}
		\centering
		\subfloat[Bunch compressor angle scan.]{
		\label{fig:peak_current_bc01}
		\includegraphics[width=7.7cm]{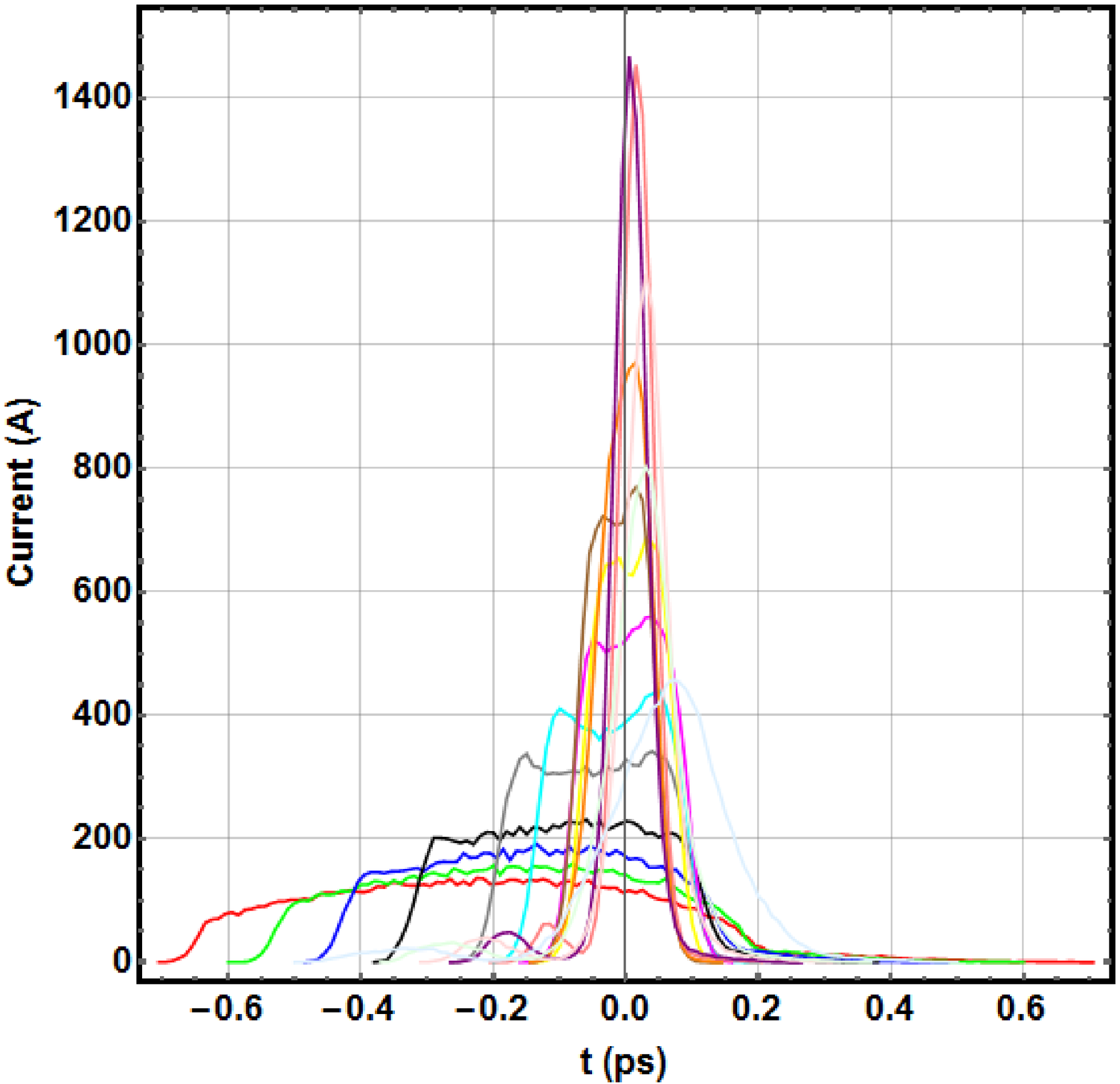}}
		\hfill
		\subfloat[Linac phase scan.]{
		\label{fig:peak_current_l01}
		\includegraphics[width=7.7cm]{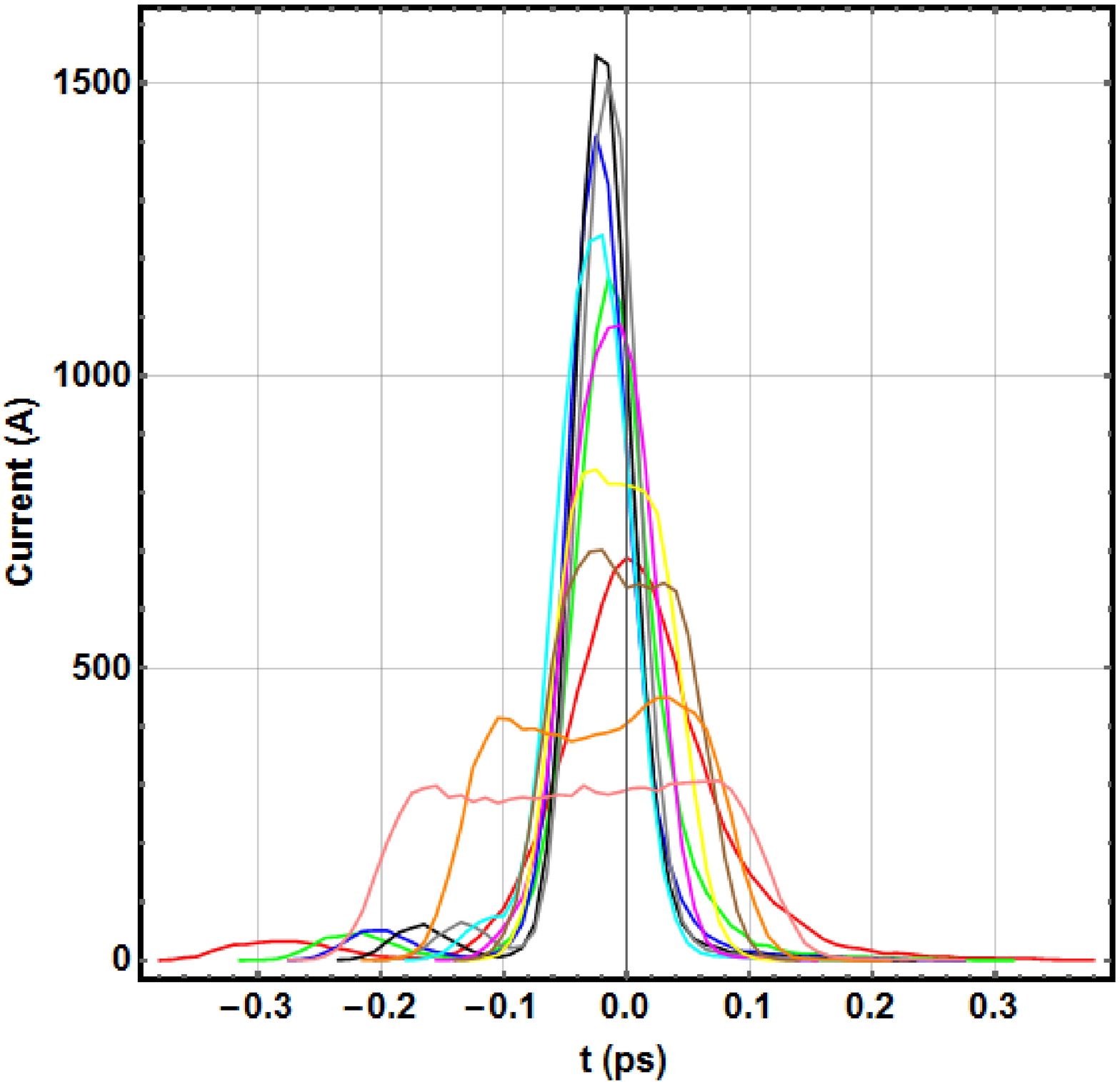}}
		\caption{\textsc{Elegant} simulation of the current profiles for compression factor scans.}
		\label{fig:peak_current}
	\end{center}
\end{figure}

The simulated current profiles for the bunch compressor and linac phase scans are shown in Fig.\,\ref{fig:peak_current}. For the linac phase scan, the largest discrepancies between the experimental data and the \textsc{CSRTrack} and \textsc{GPT} simulations occur between linac phase settings of $71.6 - 72.1$\,\si{\degree} -- in this range, the maximum current is greater than $1.3$\,\si{\kilo\ampere}. Comparing the results from simulation and experiment with $D_{par}$, we observe the most appreciable overestimation of the effect of CSR in the 1D simulation and analytic calculations when $D_{par}$ is greater than $2.5$ at any point across the chicane. When the value $D_{par}$ is smaller than this value, as during configurations with more moderate compression as in Fig.\,\ref{fig:emit_bc01}, the agreement between all of the simulation and experimental results is good.

The differences between the \textsc{Elegant} results and those from \textsc{CSRTrack} and \textsc{GPT} simulations are also noteworthy. It appears that, when the bunch undergoes maximum compression (as seen from the minimal bunch length in Figs.\,\ref{fig:emit_bc01} and \ref{fig:emit_l01}), the discrepancy between the 1D and 3D codes is largest, with \textsc{Elegant} returning an emittance value around $40$\,\si{\percent} larger than \textsc{CSRTrack}. \textsc{GPT} does return a slightly higher value for the emittance than \textsc{CSRTrack} and the experimental data around maximal compression. In order to rule out LSC in \textsc{Elegant} accounting for this difference, the parameter scans were simulated with LSC switched on and off in \textsc{Elegant}, with only a maximum reduction of $5$\,\si{\percent} in the projected emittance without LSC. Little variation was seen in the \textsc{GPT} results with space-charge switched off. Comparisons between CSR simulations and experimental data have been studied previously \cite{PhysRevAccelBeams.19.034402,PhysRevSTAB.18.030706,PhysRevSTAB.12.030704}, but only for moderate compression factors (up to around $15$ at a given bunch compressor). Indeed, at moderate compression factors -- up to around $15 - 20$, at which point the bunch length approaches $100$\,\si{\femto\second}, we see relatively good agreement between the codes and experimental data to within $10$\,\si{\percent}. It can also be seen that the \textsc{Elegant} simulations reproduce the analytic estimates for emittance growth in Figs.\,\ref{fig:emit_bc01} and \ref{fig:emit_l01}, suggesting that the code accurately reflects the predictions of the 1D theory; however, this also shows that the 1D theory may be inadequate for describing the effect of CSR in more extreme bunch compression scenarios. The fact that the codes which calculate the CSR fields directly from the retarded potentials give a closer agreement with experimental data further suggests that there are limits to the applicability of the 1D CSR approximation. 

\begin{figure}[h]
	\begin{minipage}[b]{.5\linewidth}
		\centering
		\subfloat[]{  
			\includegraphics[width=7.5cm]{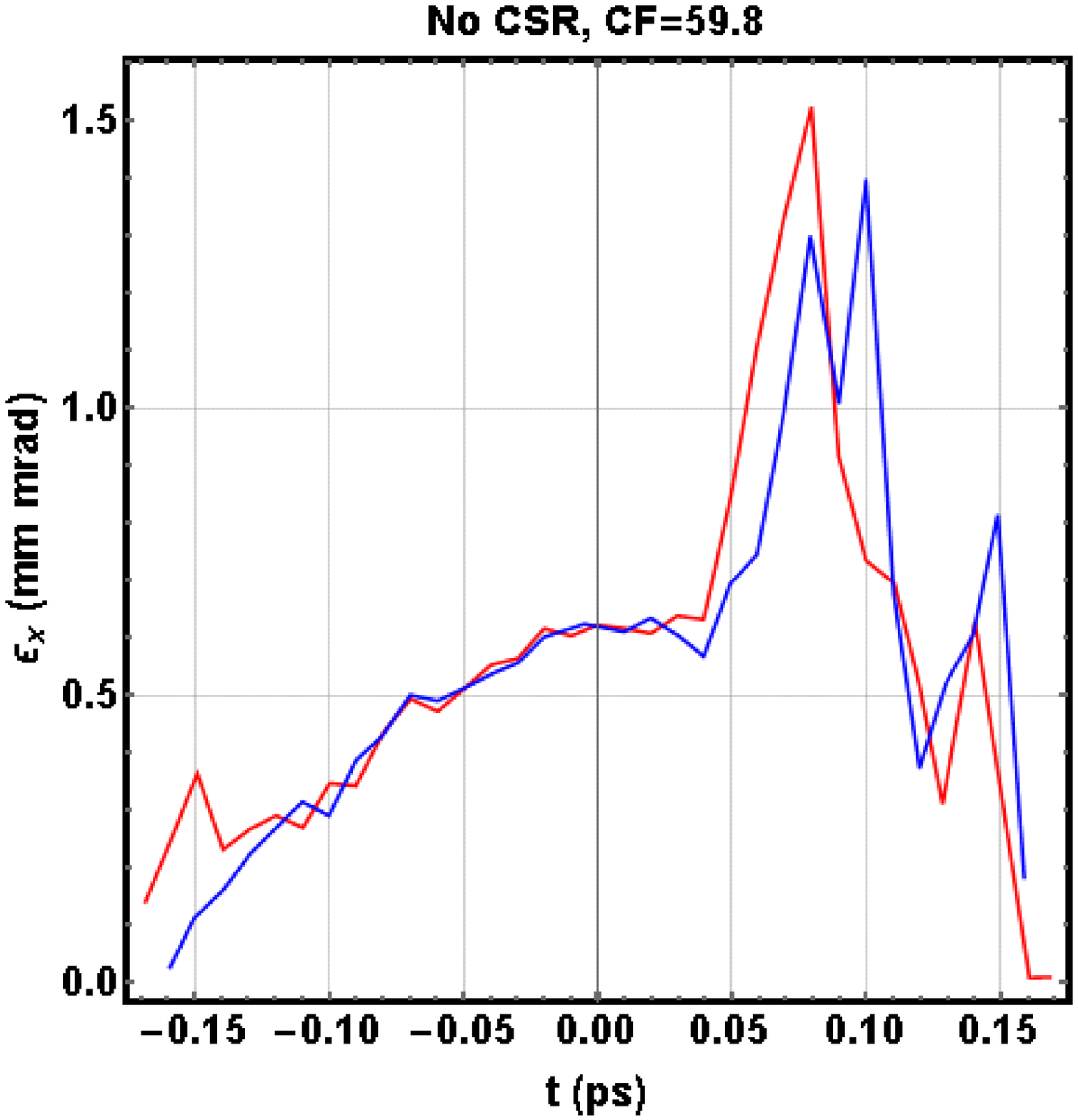} 
			\label{fig:emitplot_nocsr_bc01}}
	\end{minipage}%
	\begin{minipage}[b]{.5\linewidth}
		\centering
		\subfloat[]{  
			\includegraphics[width=7.5cm]{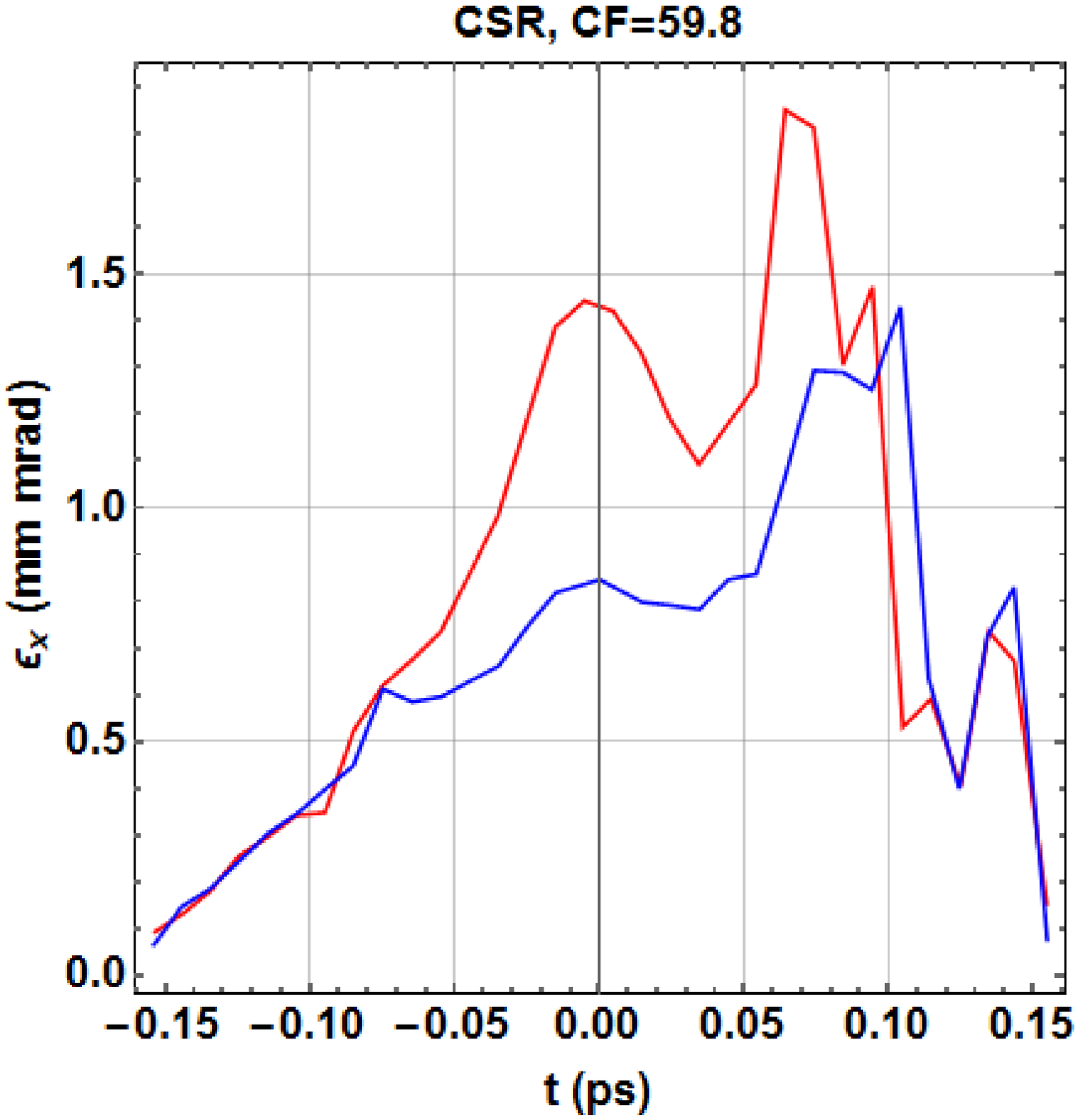} 
			\label{fig:emitplot_csr_bc01}}
	\end{minipage}
	\caption{Slice emittance for maximum compression in the bunch compressor angle scan as simulated by \textsc{Elegant} (Red) and \textsc{CSRTrack} 1D (Blue): a) with, and b) without CSR.}\label{fig:emitplots_csr_bc01}
\end{figure}

\begin{figure}[h]
	\begin{minipage}[b]{.5\linewidth}
		\centering
		\subfloat[]{  
			\includegraphics[width=7.5cm]{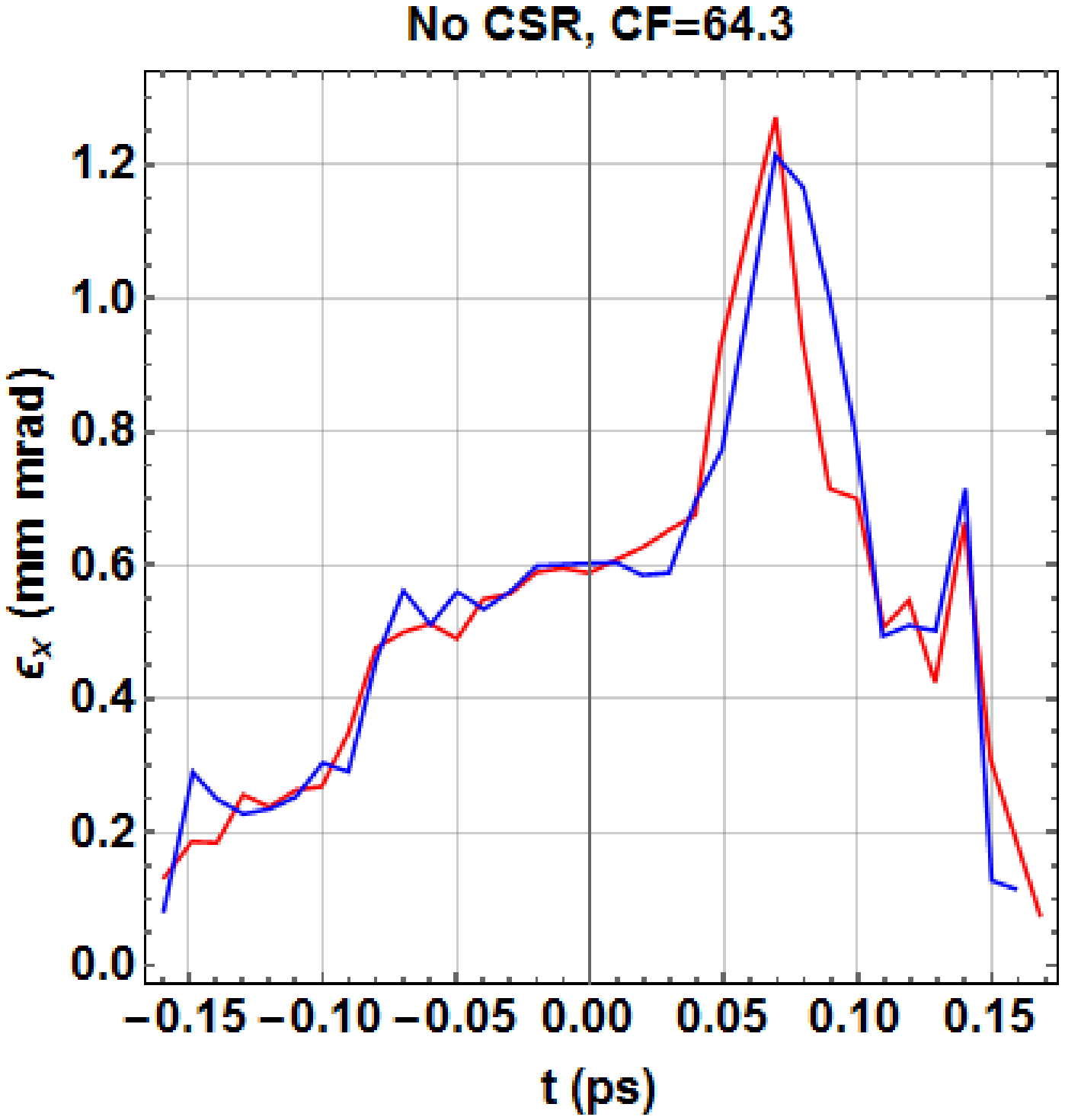} 
			\label{fig:emitplot_nocsr_l01}}
	\end{minipage}%
	\begin{minipage}[b]{.5\linewidth}
		\centering
		\subfloat[]{  
			\includegraphics[width=7.5cm]{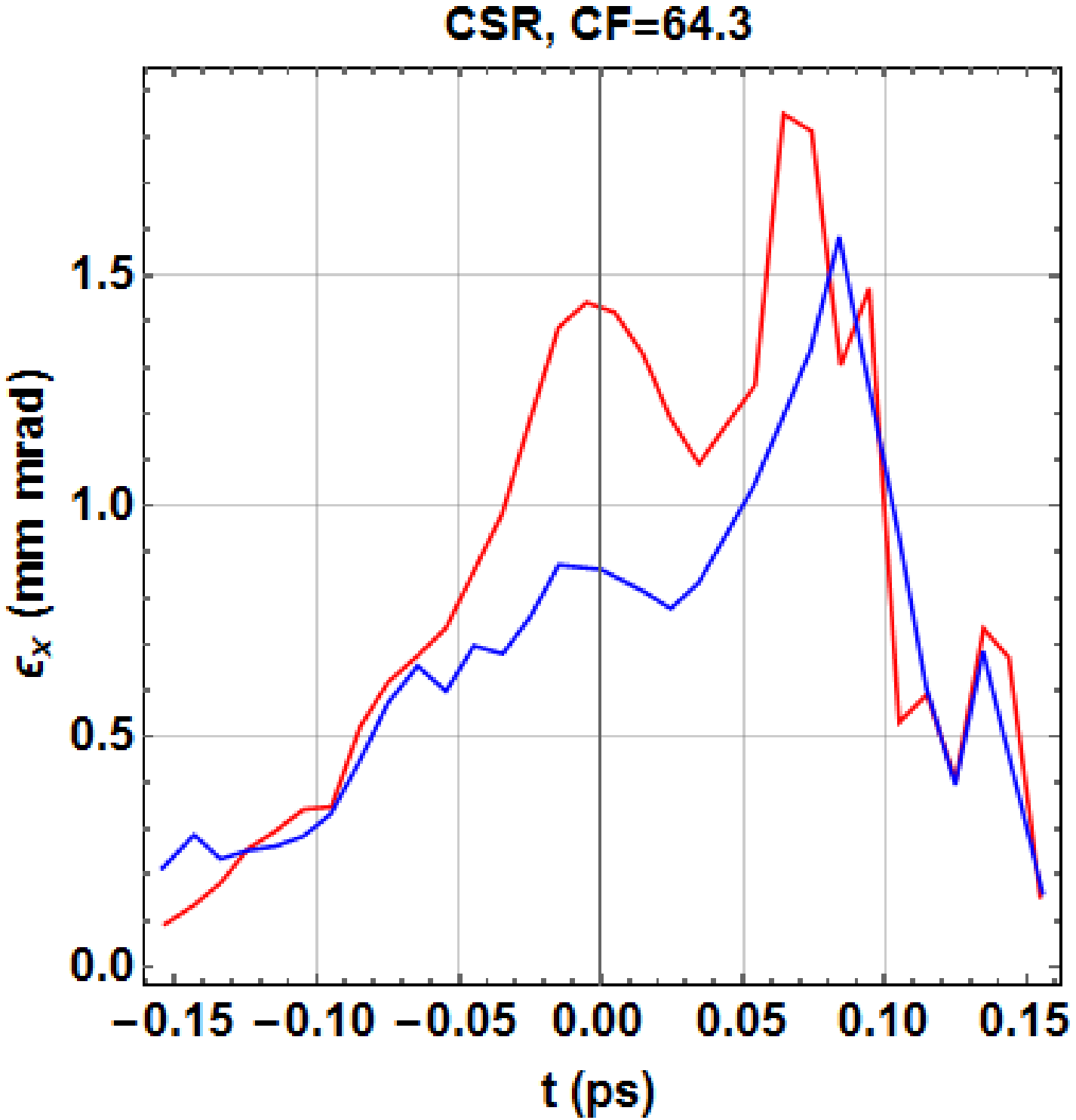} 
			\label{fig:emitplot_csr_l01}}
	\end{minipage}
	\caption{Slice emittance for maximum compression in the linac phase scan as simulated by \textsc{Elegant} (Red) and \textsc{CSRTrack} 1D (Blue): a) with, and b) without CSR.}\label{fig:emitplots_csr_l01}
\end{figure}

As the compression factor is increased -- up to a maximum value of $64$ -- more significant discrepancies between the simulation results appear. It appears that there is an overestimation of the effects of CSR in \textsc{Elegant}. By comparing the simulated slice properties for various compression factors, we can try to observe where the discrepancies arise. In order to isolate the effects of CSR, the parameter scans were run in \textsc{Elegant} and \textsc{CSRTrack} 1D with CSR switched off (see Figs.\,\ref{fig:emitplot_nocsr_bc01} and \ref{fig:emitplot_nocsr_l01}). The agreement between the codes in this case is good, and from this we can conclude that CSR \textit{is} the dominant process causing the projected emittance growth. It is also clear that the CSR-induced emittance growth is largest in the central portion of the bunch, due to the greater density of particles in this region.

Now, if the same set of parameter scans are run again with CSR switched on, (see Figs.\,\ref{fig:emitplot_csr_bc01} and \ref{fig:emitplot_csr_l01}), it can be seen that, towards maximal compression, the \textsc{Elegant} simulation returns a higher value for the horizontal slice emittance in the central portions of the bunch as compared with the results from the \textsc{CSRTrack} 1D simulation. This is the region where, for a bunch with a Gaussian longitudinal distribution, the steady-state CSR wake is largest. Slice emittance values at lower compression values (on either side of the maximum) show good agreement between the codes. The vertical emittance and current profiles are almost identical in all compression scenarios. 

\begin{figure}
\begin{center}
	\centering  
	\subfloat[Q\_L01.04 scan.]{  
	\includegraphics[width=7.5cm]{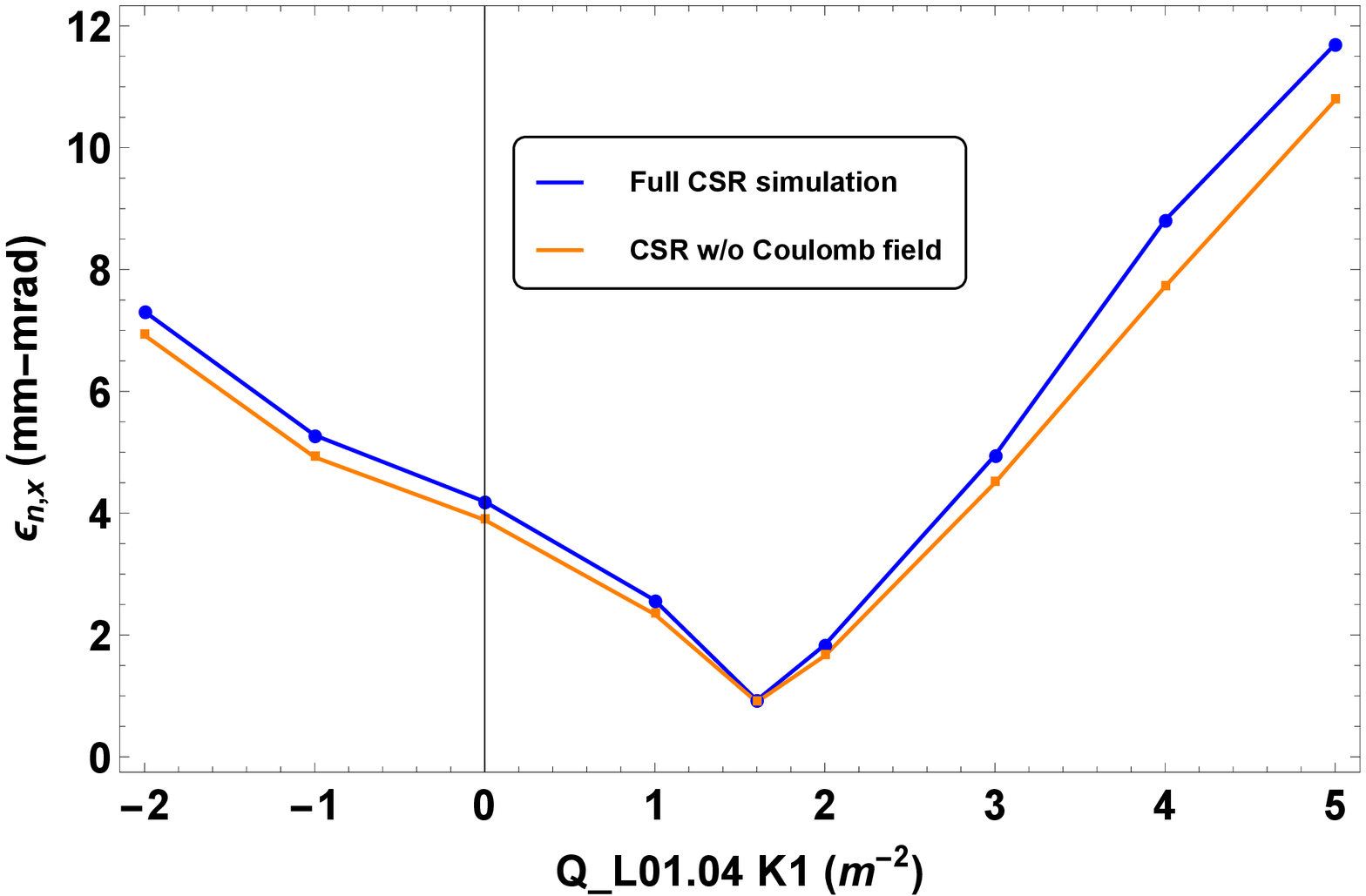} 
	\label{fig:emitplots_quad_coulomb}}         
	\hfill
	\subfloat[Linac phase scan.]{
	\includegraphics[width=7.5cm]{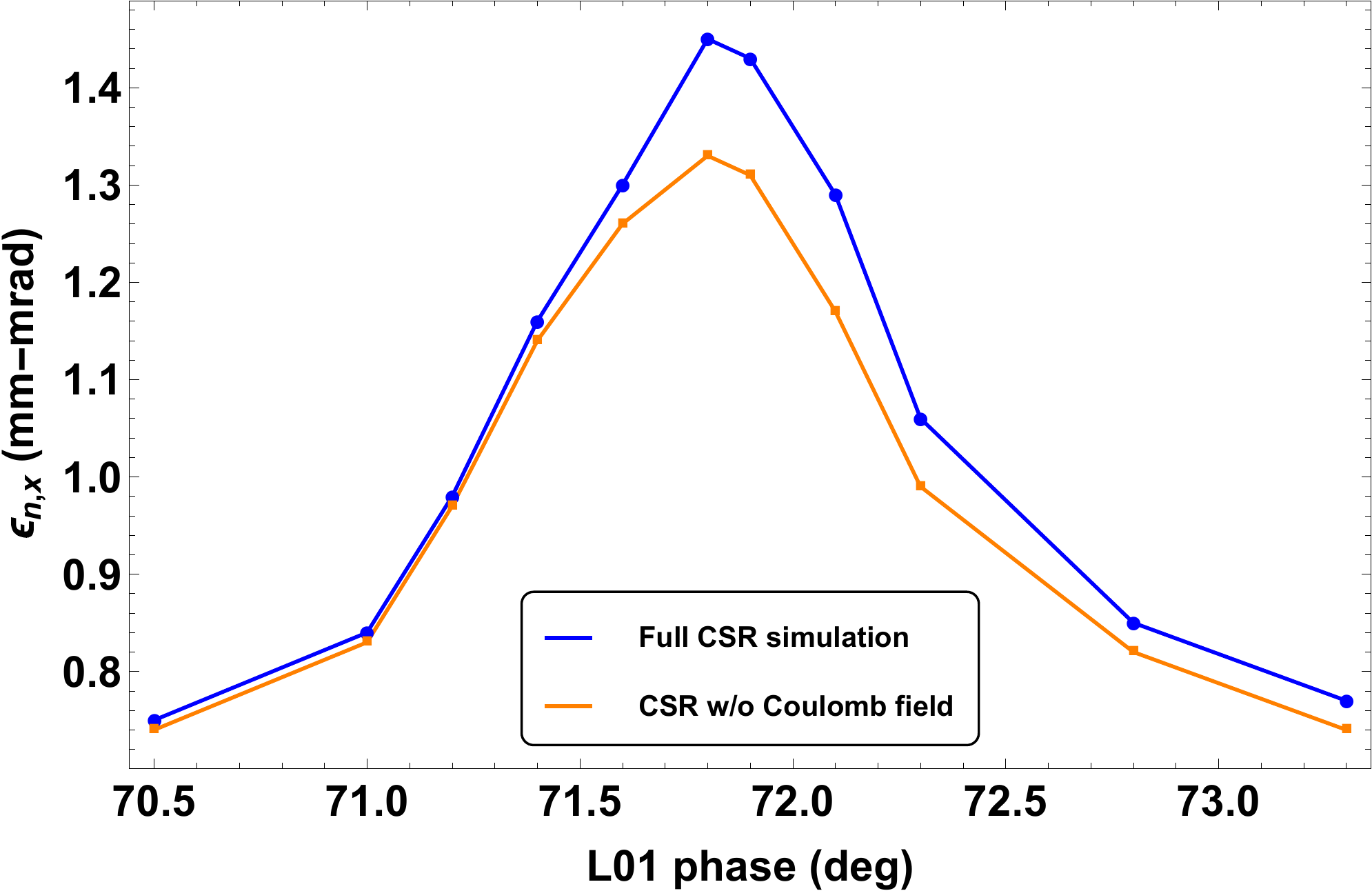}		\label{fig:emitplots_l01_coulomb}}
	\subfloat[Bunch compressor angle scan.]{
	\includegraphics[width=7.5cm]{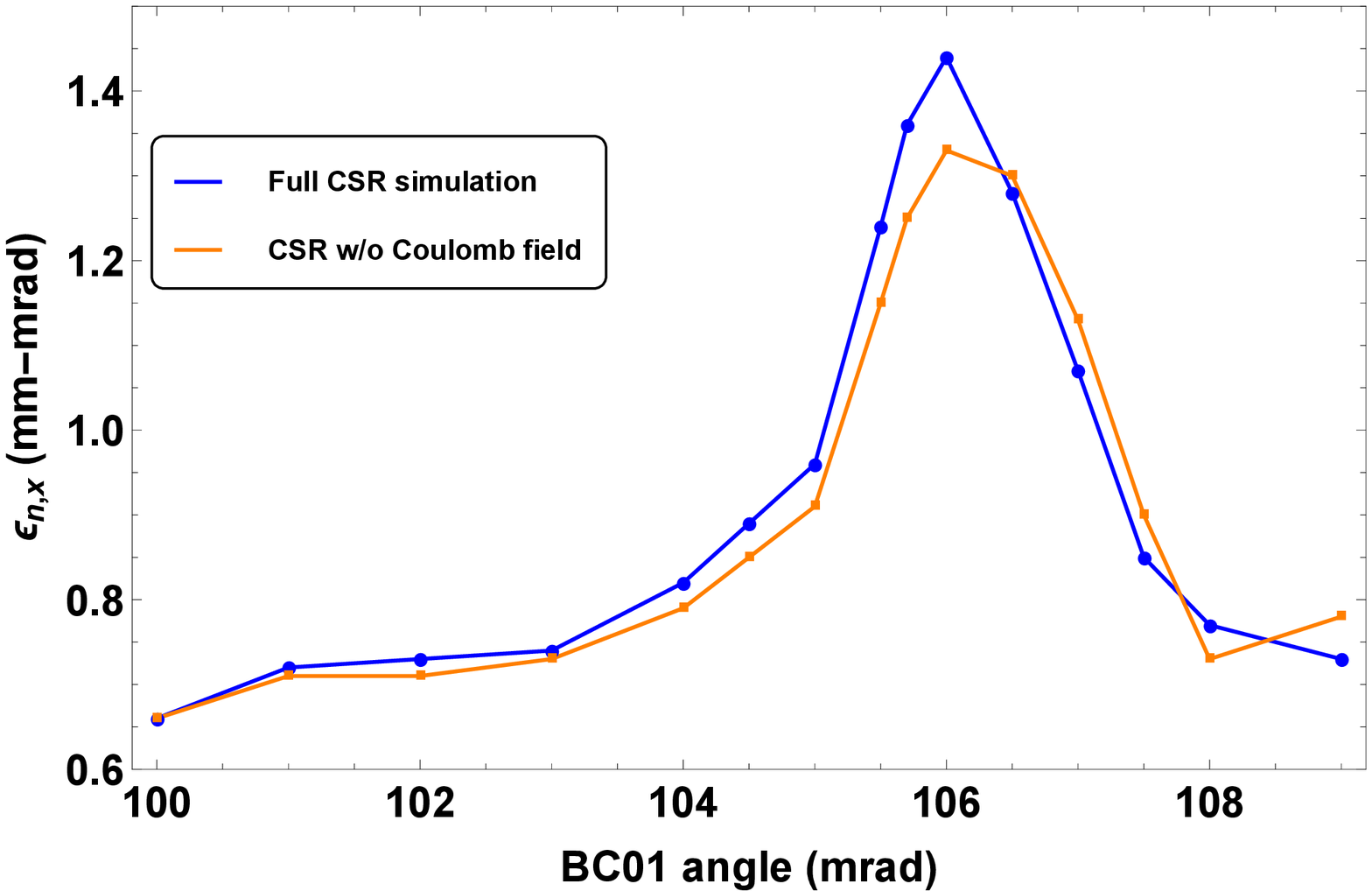}		\label{fig:emitplots_bc01_coulomb}}
	\caption{GPT simulation of emittance growth in the three scans with and without the velocity term of the Li\'enard-Wiechert field.}
	\label{fig:emitplots_coulomb}
\end{center}
\end{figure} 

Another noteworthy effect of the Li\'enard-Wiechert potentials in strong compression scenarios is revealed by the \textsc{GPT} simulation. Our results have shown that neglecting the Coulomb term across the entire compression chicane can have an effect on the final output for stronger compression factors, or for bunches with a large horizontal size. Plots comparing the final projected emittance as simulated by \textsc{GPT} for the three parameter scans are shown in Figs.\,\ref{fig:emitplots_quad_coulomb}, \ref{fig:emitplots_l01_coulomb} and \ref{fig:emitplots_bc01_coulomb}, with the Coulomb term switched on and off. It can be seen that, approaching maximal compression, or largest transverse bunch size, the Coulomb term can have a significant impact on the final CSR-induced emittance growth. In the case of a larger bending angle in the chicane, this can be understood as the catch-up distance for the Coulomb term being shorter, and similarly for a bunch with minimal chirp around the linac phase for maximal compression. When the bunch has a larger transverse size due to the focusing into the chicane, the radiating cone is also larger, and so the Coulomb interaction between the tail and head of the bunch will also make a larger contribution. It is also likely that this neglected term will have a more significant impact for accelerator configurations with a higher density of bends, such as in arcs in energy recovery linacs (ERLs). These results give further evidence of the importance of correctly simulating CSR effects in dispersive regions.

\section{Conclusions}\label{sec:conclusions}

We have found that the longitudinal electric field as observed in the CSR interaction before the entrance to, and after the exit of, a bending magnet has a qualitatively different behaviour than is commonly assumed. In particular, the contribution to the CSR field from the Coulomb field of the Li\'enard-Wiechert potential cannot be neglected when calculating entrance transient effects, and in order to correctly model this interaction, these fields must be taken into account. The observations of this paper are interesting technologically because they suggest that it may be possible to design an optimized magnet (or system of magnets) in which the CSR impact of the magnet itself is partially cancelled by that of the drift directly after it, thereby reducing adverse effects like emittance growth and microbunching gain, particularly in more complex transport systems such as compressive arcs in ERLs.

We have also detailed a comparison between experimental measurements and simulation results to determine the effect of CSR on projected emittance growth, and have shown some agreement with \textsc{Elegant}, \textsc{GPT} and \textsc{CSRTrack} simulations. Good agreement between the simulations and FERMI measurements is seen when the compression, rf parameters, and matching are closest to nominal. As the compression ratio increases, the differences between the simulation methods become more clear. As expected, the 1D simulation results diverge more significantly both from the experimental data, and from simulations results with codes which take the transverse extent of the bunch into account in situations where the Derbenev criterion is strongly violated, and the 1D CSR approximation breaks down. This study has also shown the importance of correctly accounting for the Li\'enard-Wiechert interaction across the entirety of a dispersive region, and for both the region that precedes it and that which comes after. 

\section*{Acknowledgements}

The authors would like to acknowledge the support of the Industrial Liason Office of Elettra Sincrotrone Trieste. A.D.B. would also like to thank Andy Wolski and Bruno Muratori for their support and advice.

\appendix 

\section{Derivation of CSR Entrance Transient}\label{sec:appendix_entrance}

We begin by calculating a number of distances shown in Fig.\,\ref{fig:csr_entrance_transient}. The angle $\phi$ is the angle between the receiving particle at position $\vec{r_1}$ and the entrance of the arc. The two orthogonal directions of this arc $(h,D)$ are given by $h=2R\sin^2(\phi/2)$ and $D=R\sin(\phi)$, and therefore the distance $\rho$ between the emitter at the retarded position $\vec{r_0}'$, at a distance $y$ before the entrance to the magnet, and the receiver at $\vec{r_1}$ is:

\begin{equation}
\rho = \sqrt{h^2 + (y + D)^2} = \sqrt{4R^2\sin^2(\phi/2) + 2Ry\sin(\psi) + y^2}.
\end{equation}

The time taken by electromagnetic signals to travel from emitter to observer is $t - t' = \rho/c$. During this same time, the bunch must have traveled a distance $R\phi + y - \Delta z$ along the path in order to have the observing electron at the position sketched in Fig.\,\ref{fig:csr_entrance_transient} at time $t$, where $\Delta z = z - z'$ is the instantaneous distance between both electrons. Therefore $t - t' = (R\phi + y - \Delta z)/(\beta c)$, from which follows the retardation condition:

\begin{equation} \label{eq:retardationcondition}
\Delta z = y + R\phi - \beta \rho.
\end{equation}

Two more useful lengths sketched in Fig.\,\ref{fig:csr_entrance_transient} are:

\begin{equation}
w = \frac{y}{y + D}h = \frac{2yR\sin^2(\phi/2)}{y + R\sin(\phi)}
\end{equation}

and

\begin{equation}
L = \frac{D}{y + D}\rho = \frac{R\sin(\phi)}{y + R\sin(\phi)}\rho,
\end{equation}

\noindent where $L$ is the distance between the entrance to the magnet and the observation point $\vec{r_1}$. These lengths can be used to derive the cosine and sine of the angles $\xi$ and $\theta$ between the vectors $\vec{n}$, $\vec{\beta}$ and $\vec{\beta}'$ which are required to evaluate Eq.\,\ref{eq:lwfield}. The triangle defined by the emitter and the endpoints of $h$ gives:

\begin{equation}
\cos(\theta) = \frac{y + D}{\rho} = \frac{y + R\sin(\phi)}{\rho}
\end{equation}

\noindent and

\begin{equation}
\sin(\theta) = \frac{h}{\rho} = \frac{2R\sin^2(\phi / 2)}{\rho}.
\end{equation}

In order to calculate $\xi$ we need to use its complementary angle $\eta$. Using the cosine and sine rules on the triangle defined by $\eta$ and $\phi$ gives:

\begin{equation}
\cos(\xi) = \sin(\eta) = \frac{R - w}{L}\sin(\phi) = \frac{R\sin(\phi) + y\cos(\phi)}{\rho}
\end{equation}

\noindent and

\begin{equation}
\sin{\xi} = \cos{\eta} = \frac{R^2 + L^2 - (R - w)^2}{2RL} = \frac{2\sin(\phi/2) (R\sin(\phi/2) + y\cos(\phi/2))}{\rho}.
\end{equation}

Having these angles available, we can now calculate the point-to-point Li\'enard-Wiechert field of the emitter $\vec{E}^{PP}$ at the position of the receiver. Since we require only the parallel component, we can take the inner product $\vec{\beta}\cdot\vec{E}^{PP}$. Additionally, because the emitter is in uniform motion, its retarded electric field is given only by the velocity term of the Li\'enard-Wiechert field, yielding:

\begin{equation}\label{eq:lwvelocityfield}
E_{||}^{PP, ent} = \frac{e}{4\pi\epsilon_0 \gamma^2} \frac{\vec{n}\cdot\vec{\beta} - \vec{\beta}'\cdot\vec{\beta}}{(1 - \vec{n}\cdot\vec{\beta})^3 \rho^2}.
\end{equation}

\noindent These inner products can be expressed in terms of the angle $\phi$ as follows:

\begin{equation}\label{eq:nbeta}
\vec{n}\cdot\vec{\beta} = \beta\cos(\xi) = \beta\frac{R\sin(\phi) + y\cos(\phi)}{\rho},
\end{equation}

\begin{equation}\label{eq:nbetaprime}
\vec{n}\cdot\vec{\beta}' = \beta\cos{\theta} = \beta\frac{y + R\sin(\phi)}{\rho}
\end{equation}

and

\begin{equation}\label{eq:betabetaprime}
\vec{\beta}\cdot\vec{\beta}' = \beta^2 \cos(\phi).
\end{equation}

\noindent Substituting into Eq.\,\ref{eq:lwvelocityfield} gives:

\begin{equation}
E_{||}^{PP, ent}(y) = \frac{e}{4\pi\epsilon_0 \gamma^2}\frac{(y - \beta\rho) \cos(\phi) + R\sin(\phi)}{\left(\rho - \beta(y + R\sin(\phi))\right)^3}.
\end{equation}

This is the field observed by a single point particle at an angle $\phi$ into the arc, produced by another single point particle at a distance $y$ before the entrance of the arc. In order to obtain the field due to a bunch of particles, the bunch with a charge density $Ne\lambda(z)$ should be thought of as a number of point particles at positions $s = s_c + z$, each with charge $Ne\lambda(z)dz$, where $s$ is the absolute position along the path, $z$ is the position within the bunch relative to the bunch centroid and $s_c$ the position of the centroid. Summing up the contributions from all the fields of these point particles gives:

\begin{equation}\label{eq:totalcsrentrancefield}
E_{||}^{ent}(z, y) = Ne\int_{-\infty}^{z_{max}} E_{||}^{PP, ent}\left(y(z')\right)\lambda(z')dz' + E_{||}^{SS},
\end{equation}

\noindent where the first term represents the field contribution at the position of the receiver due to the part of the bunch that is still before the magnet entrance at the time of emission, $E_{||}^{SS}$ is the contribution to the field due to the part of the bunch that is inside the magnet at the same time, and $z_{max}$ is the position in the bunch giving the boundary between these two parts. In order to evaluate Eq.\,\ref{eq:totalcsrentrancefield} directly, an explicit relation between the current position of the emitter $z'$ and its position at the time of emission $y$ is required. This can be done by changing the integration variable from $z'$ to $y$, and so we can use Eq.\,\ref{eq:retardationcondition} to calculate $dz'/dy$:

\begin{equation}
\frac{dz'}{dy} = -\frac{\rho - \beta(y + R\sin(\phi))}{\rho}.
\end{equation}

\noindent Given this relation and that from Eq.\,\ref{eq:steadystateapprox}, we have an equation for $E_{||}^{SS}$, and the total CSR entrance field thus becomes:

\begin{multline}
E_{||}^{ent}(z, y) = Ne \int_{-\infty}^{z_{max}} E_{||}^{PP, ent}(y)\lambda(z'(y))\frac{dz'}{dy}dy + E_{||}^{SS} \\
= E_{||}^{SS} + \frac{Ne}{4\pi\epsilon_0 \gamma^2}\int_{0}^{d}\frac{(y - \beta \rho(y))\cos(\phi) + R\sin(\phi)}{(\rho(y) - \beta(y + R\sin(\phi)))^2 \rho(y)} \lambda(z - \Delta(y))dy,
\end{multline}

\noindent where $\Delta(y) = y + R\phi - \beta\rho(y)$, and $d$ the length of the drift before the magnet taken into account for the calculation of the CSR field. The upper integration boundary of this expression arises due to the finite length of the straight section before the entrance to the magnet. 

\section{Derivation of CSR Exit Transient}\label{sec:appendix_exit}

This derivation will parallel that given above in \ref{sec:appendix_entrance} for the entrance transients. To evaluate Eq.\,\ref{eq:lwfield}, we first calculate a number of lengths indicated in Fig.\,\ref{fig:csr_exit_transient}. The angle $\psi$ is the angle between the emitter and the end of the arc. The lengths $(h,D)$ along two orthogonal directions associated with this arc are given by $h=2R\sin^2(\psi/2)$ and $D=R\sin(\psi)$. Therefore $\rho$ is equal to:

\begin{equation}\label{eq:rho}
\rho = \sqrt{h^2 + (x + D)^2} = \sqrt{4R^2\sin^2(\psi/2) + 2Rx\sin(\psi) + x^2},
\end{equation}

\noindent where $x$ is the distance from the exit edge of the magnet to the observing electron. The time taken by electromagnetic signals to travel from emitter to observer is $t - t' = \rho/c$. During this same time, the bunch must have traveled a distance $R\psi + x - \Delta z$ along the path in order to have the observing electron at the position sketched in Fig.\,\ref{fig:csr_exit_transient} at time $t$, where $\Delta z = z - z'$ is the instantaneous distance between both electrons. Therefore $t - t' = (R\psi + x - \Delta z)/(\beta c)$, from which follows the retardation condition:

\begin{equation} \label{eq:retardationcondition1}
\Delta z = x + R\psi - \beta \rho.
\end{equation}

\noindent Two more useful lengths sketched in Fig.\,\ref{fig:csr_exit_transient} are:

\begin{equation}
w = \frac{2xR\sin^2(\psi/2)}{(x + R\sin(\psi))}
\end{equation}

and

\begin{equation}
L=\frac{R\rho \sin(\psi)}{(x + R\sin\psi)},
\end{equation}

\noindent where $L$ is the distance between the emitting electron at $\vec{r'}$ and the exit of the magnet. These lengths can be used to derive the angles $\xi$ and $\theta$ between the three vectors $\vec{n},\vec{\beta}$ and $\vec{\beta}'$ indicated in Fig.\,\ref{fig:csr_exit_transient}. The triangle defined by $\vec{r}$ and the endpoints of $h$ gives $\cos(\xi) = (x + R\sin(\psi))/\rho$ and	$\sin(\xi) = (2R/\rho)\sin^2(\psi/2)$. Using the cosine and sine rules on the triangle defined by $\eta$ and $\psi$ gives:

\begin{equation}
\cos \theta = (R\sin(\psi) + x\cos(\psi))/\rho
\end{equation}
and 

\begin{equation}
\sin \theta = 2\sin(\psi/2) (R\sin(\psi/2) + x\cos(\psi/2))/\rho.
\end{equation}

Since we are interested only in the component of the field parallel to the direction of $\vec{\beta}$, we take the inner product $\vec{\beta} \cdot \vec{E}$. As it turns out, both the radiation term and the velocity term of the Li\'enard-Wiechert field make a significant contribution, even in the ultrarelativistic limit. Expanding the triple vector product in the radiation term of the field (Eq.\,\ref{eq:lwfield}) and taking the inner product of the full Li\'enard-Wiechert field with $\vec{\beta}$ gives:

\begin{equation} \label{eq:lwexpandedfield}
E_{||}^{PP} = \frac{\vec{\beta} \cdot \vec{E}}{\beta} = \frac{e}{4\pi \epsilon_0 \beta}\left( \frac{\left(\vec{n}.\vec{\beta}-\vec{\beta}\cdot \vec{\beta}'\right)}{\gamma^2\left(1-\vec{n}\cdot \vec{\beta}'\right)^3 \rho^2} \frac{\left(\vec{n} \cdot \vec{\beta} - \vec{\beta} \cdot \vec{\beta}'\right) \left(\vec{n} \cdot \dot{\vec{\beta}}' \right) - \left(1 - \vec{n} \cdot \vec{\beta}' \right)\left(\vec{\beta} \cdot \dot{\vec{\beta}}'\right)}{c\left(1-\vec{n}\cdot \vec{\beta}'\right)^3 \rho}\right).
\end{equation} 

\noindent The superscript $PP$ indicates that Eq.\,\ref{eq:lwexpandedfield} gives the field of a point particle. We can calculate the inner products in this expression as follows:

\begin{equation}
\vec{n} \cdot \vec{\beta} = \beta \cos(\xi) = \beta \frac{x + R\sin(\psi)}{\rho},
\end{equation}

\begin{equation}
\vec{n} \cdot \vec{\beta}' = \beta \cos(\theta) = \beta \frac{R\sin(\psi) + x\cos(\psi)}{\rho},
\end{equation}

\begin{equation}
\vec{n} \cdot \dot{\vec{\beta}}' = \frac{\beta^2 c}{R}\sin(\theta) = \frac{\beta^2 c}{R} \frac{2\sin\left(\psi/2\right)\left(R\sin\left(\psi/2\right) + x\cos\left(\psi/2\right)\right)}{\rho},
\end{equation}

\begin{equation}
\vec{\beta} \cdot \vec{\beta}' = \beta^2 \cos(\psi),
\end{equation}

\begin{equation}
\vec{\beta} \cdot \dot{\vec{\beta}}' = \frac{\beta^3 c}{R} \sin(\psi).
\end{equation}

\noindent Substituting these expressions into Eq.\,\ref{eq:lwexpandedfield} and separating the first and second terms into the velocity and radiation components, respectively, results in the single-particle longitudinal components of the CSR field:

\begin{equation} \label{eq:singleparticlecsrvelocityfield}
E_{||, vel}^{PP,exit}(z, x) = \frac{e \beta}{4 \pi \epsilon_0 \gamma^2}\frac{x + R\sin(\psi) - \beta \rho \cos(\psi)}{\left(\rho - \beta \left(R\sin(\psi) + x\cos(\psi) \right) \right)^3},
\end{equation}

\begin{multline} \label{eq:singleparticlecsrfield}
E_{||, rad}^{PP,exit}(z, x) = \frac{e \beta^2}{4 \pi \epsilon_0 R}\left(\frac{2\sin(\psi/2)\left(x + R\sin(\psi) - \beta \rho \cos(\psi)\right) \left(R\sin(\psi/2) + x\cos(\psi/2)\right)}{\left(\rho - \beta \left(R\sin(\psi) + x\cos(\psi) \right) \right)^3} \right. \\
\left. - \frac{\rho\sin(\psi)}{\left(\rho - \beta\left(R\sin(\psi) + x\cos(\psi)\right) \right)^2}\right).
\end{multline}

\noindent These are the contributions of the velocity and radiation terms to the field observed by a single electron at distance $x$ after the exit of the arc, produced by another single electron at angle $\psi$ before the exit of the arc. Which particular electron of the bunch distribution actually is at angle $\psi$ at the required time of emission is governed by the retardation condition Eq.\,\ref{eq:retardationcondition1}. One may be tempted to neglect the velocity term $E_{||, vel}^{PP}$ in the ultrarelativistic limit on account of the factor $\gamma^{-2}$. However, the radiation term contains an additional small factor $\sin\left(\psi/2\right)$ in the numerator, and so in the end both terms are comparable in size. The combined field $E$ due to all electrons between the tail of the bunch and the observing electron is obtained by adding the fields $E^{PP}$ of the individual particles. This results in:

\begin{equation}\label{eq:totalcsrexitfield}
E_{||}^{exit}(z, x) = Ne\int_{-\infty}^{z} E_{||}^{PP,exit}\left(\psi(z')\right)\lambda(z')dz',
\end{equation}

\noindent where $N$ is the number of particles in the bunch and $\lambda(z')$ is the charge distribution normalised such that $\int_{-\infty}^{+\infty}\lambda(z')dz' = 1$. To evaluate this integral directly, an explicit relation $\psi(z')$ between the current position of the emitter $z'$ and the position at time of emission given by $\psi$ is necessary. To avoid this complication, we change the integration variable from $z'$ to $\psi$. This requires the derivative $dz'/d\psi$, which from Eq.\,\ref{eq:retardationcondition1} is $-R\left(1 - \beta\left(R\sin(\psi) + x\cos(\psi)\right)/\rho\right)$. Eq.\,\ref{eq:totalcsrexitfield} thus becomes:

\begin{equation}
E_{||}^{exit}(z, x) = Ne\int_{0}^{\phi_m} E_{||}^{PP, exit}(\psi) \lambda\left(z'(\psi)\right) \frac{dz'}{d\psi}d\psi = E_{||, vel}^{exit}(z, x) + E_{||, rad}^{exit}(z, x),
\end{equation}

\noindent and the velocity and radiation terms are defined as:

\begin{equation}\label{eq:totalcsrfieldvelocityexita}
E_{||, vel}^{exit}(z, x) = \frac{N e \beta R}{4\pi \epsilon_0 \gamma^2} \int_{0}^{\phi_m} \frac{x + R\sin(\psi) - \beta \rho \cos(\psi)}{\left(\rho - \beta \left(R \sin(\psi) + x\cos(\psi)\right) \right)^2 \rho} \lambda\left(z'(\psi)\right) d\psi,
\end{equation}

and

\begin{multline}\label{eq:totalcsrfieldradiationexita}
E_{||, rad}^{exit}(z, x) = \frac{Ne \beta^2}{4\pi \epsilon_0}\int_{0}^{\phi_m}\left( \frac{2\sin(\psi/2)\left(x + R\sin(\psi) - \beta \rho \cos(\psi)\right)\left(R\sin(\psi/2) + x\cos(\psi/2)\right)}{\left(\rho - \beta\left(R\sin(\psi) + x\cos(\psi)\right)\right)^2\rho} \right. \\
\left. - \frac{\sin(\psi)}{\rho - \beta\left(R\sin(\psi) + x\cos(\psi)\right)}\right) \lambda\left(z'(\psi)\right)d\psi.
\end{multline}

In this expression, $x = x_c + z$ is the position of the evaluation point $s$ with respect to the exit edge of the magnet, with $x_c$ the distance from exit edge to bunch centroid and $z$ the position in the bunch where the field is evaluated relative to the bunch centroid. In the integral, the charge density should be evaluated at $z'$, which from Eq.\,\ref{eq:retardationcondition} is given by $z'(\psi) = x_c + z - x_c - z' = -x_c - R\psi + \beta\rho$.

The expressions for $E_{||, vel}^{exit}$ and $E_{||, rad}^{exit}$ give the Li\'enard-Wiechert field of a bunch exiting a circular arc, without any ultrarelativistic or small-angle approximations. However, through applying these approximations we can arrive at a simpler form for these fields. Given that the integrands in these expressions are strongly peaked around a small range of $\psi \ll 1$, approximations can be made using Taylor expansions -- although it should be noted that the approximation $\psi \ll x/R$ cannot be used, as the post-bend distance $x$ may also be small. First, we re-evaluate the distance $\rho$ (Eq.\,\ref{eq:rho}) between the emitter at the retarded time and the receiver at the current time:

\begin{multline}
\rho \approx \sqrt{4R^2\left(\frac{1}{4} \psi^2 - \frac{1}{48} \psi^4\right) + 2Rx\left(\psi - \frac{1}{6} \psi^3\right) + x^2} = R(\psi + x_n)\sqrt{1 - \frac{\frac{1}{12} \psi^4 + \frac{1}{3} x_n \psi^3}{(\psi + x_n)^2}} \\
\approx R\left(\psi + x_n - \frac{\psi^2}{24}\frac{\psi^2 + 4\psi x_n}{\psi + x_n}\right),
\end{multline}

\noindent where $x_n = x/R$. This approximation can now be applied to Eqs.\,\ref{eq:totalcsrfieldvelocityexita} and \ref{eq:totalcsrfieldradiationexita}. Expanding all the trigonometric functions results in:

\begin{equation}\label{eq:totalcsrexitwakevelocitytaylor}
E_{||, vel}^{exit}(z, x) \approx \frac{8 Ne}{3\pi \epsilon_0} \int_{0}^{\phi_m} \frac{\gamma^{-2} N_1(\psi) + \psi^2 N_2(\psi) + ...}{\left(\gamma^{-2} D_1(\psi) + \psi^2 D_2(\psi)\right)^2}\gamma^{-2} \lambda\left(z'(\psi)\right)d\psi,
\end{equation}

\begin{equation}\label{eq:totalcsrexitwakeradiationtaylor}
E_{||, vel}^{exit}(z, x) \approx \frac{Ne}{\pi \epsilon_0} \int_{0}^{\phi_m} \frac{\gamma^{-2} N_3(\psi) + \psi^2 N_4(\psi) + ...}{\left(\gamma^{-2} D_1(\psi) + \psi^2 D_2(\psi)\right)^2}\psi^2 \lambda\left(z'(\psi)\right)d\psi,
\end{equation}

\noindent where (to quadratic order in $\psi$ and $x_n$):

\begin{subequations}
\begin{gather}
	N_1(\psi) = 3\left(\psi + x_n\right)^2 + ... \\
	N_2(\psi) = \left(\psi + x_n\right) \left(2\psi + 3x_n\right) + ... \\
	N_3(\psi) = 4\left(\psi + x_n\right)^2 + ... \\
	N_4(\psi) = \left(\psi + 2 x_n\right)^2 + ... \\
	D_1(\psi) = 4\left(\psi + x_n\right)^2 + ... \\
	D_2(\psi) = \left(\psi + 2x_n\right)^2 + ...
\end{gather}
\end{subequations}

There are two important results in Eqs.\,\ref{eq:totalcsrexitwakevelocitytaylor} and \ref{eq:totalcsrexitwakeradiationtaylor}. Firstly, the velocity field is suppressed by a factor $\gamma^{-2}$, so in many cases this field is negligible with respect to the radiation field. However, the integrand of the radiation field also contains the factor $\psi^2$, which suppresses this field at small angles. Therefore, there is a small part of the integration interval $\psi \lesssim \gamma^{-1}$ in which the velocity field dominates, rather than the radiation field, and vice versa for the regime $\psi \gtrsim \gamma^{-1}$. Secondly, in both the velocity field and the radiation field individually, the same two regimes can be distinguished when considering the importance of the terms proportional to $\gamma^{-2}$ against those proportional to $\psi^2$. Therefore, in the regime where the velocity field dominates, the significant terms are $N_1$ and $D_1$, and in the regime with the dominant radiation field, the significant terms are $N_4$ and $D_2$. Combining these results, we can make a further approximation:

\begin{equation}\label{eq:totalcsrfieldexitapprox}
E_{||}^{exit}(z, x) \approx \frac{8 Ne}{3\pi \epsilon_0} \int_{0}^{\phi_m} \frac{N_1(\psi)}{D_1(\psi)^2} \lambda\left(z'(\psi)\right)d \psi + \frac{Ne}{\pi \epsilon_0} \int_{0}^{\phi_m} \frac{N_4(\psi)}{D_2(\psi)^2}\lambda\left(z'(\psi)\right)d\psi.
\end{equation}

\noindent An approximation to $dz/d\psi$ (from Eq.\,\ref{eq:retardationcondition}) can be made in the ultrarelativistic limit, as $\gamma \to \infty$:

\begin{equation}
\frac{dz'}{d\psi} = -\frac{R}{\rho}\left(\rho - \beta\left(R\sin(\psi) + x\cos(\psi)\right)\right) \approx -\frac{R\psi^2 \left(\psi + 2x_n\right)^2}{8\left(\psi + x_n\right)^2}.
\end{equation}

\noindent Given that the velocity term is only significant in a small range $\psi \lesssim \gamma^{-1} \ll 1$ for sufficiently high beam energy, and that over this range the charge density will not vary significantly, we can approximate the field as follows:

\begin{equation}
E_{||, vel}^{exit}(z, x) \approx \frac{8 N e}{3\pi \epsilon_0} \int_{0}^{\phi_m} \frac{N_1 (\psi)}{D_1 (\psi)^2} \lambda(z'(\psi))d\psi \approx \frac{N e}{2\pi \epsilon_0} \lambda(z'(0)) \int_{0}^{\phi_m} \frac{d\psi}{\left(\psi + x_n \right)^2} = \frac{N e}{2\pi \epsilon_0} \frac{\lambda(z'(0))}{x_n}.
\end{equation}

Integrating the second term on the right-hand side of Eq.\,\ref{eq:totalcsrfieldexitapprox} (the radiation component) by parts gives:

\begin{equation}
E_{||, rad}^{exit}(z, x) \approx \frac{Ne}{\pi \epsilon_0}\left(\frac{\lambda (z - \Delta z_{max})}{\phi_m R + 2x_n} - \frac{\lambda(z)}{2 x_n} + \int_{z - \Delta z_{max}}^{z - \Delta z_{min}} \frac{\partial \lambda (z')}{\partial z'} \frac{dz'}{\psi(z')R + 2x_n}\right).
\end{equation}

\noindent In the integrand of this expression, $\psi(z)$ is defined implicitly by the relation:

\begin{equation}
z - z' = f(\psi) = \frac{R \psi^3}{24}\frac{R \psi + 4x}{R \psi + x}.
\end{equation}

\noindent and $\Delta z_{max} = f(\phi_m)$. Now, it can be seen that there is an exact cancellation between the second term of $E_{||, rad}^{exit}$ and $E_{||, vel}^{exit}$, leading to a final approximation for the exit transient field:

\begin{equation}
E_{||}^{exit}(z, x) \approx \frac{Ne}{\pi \epsilon_0}\left(\frac{\lambda (z - \Delta z_{max})}{\phi_m R + 2x_n} + \int_{z - \Delta z_{max}}^{z} \frac{\partial \lambda (z')}{\partial z'} \frac{dz'}{\psi(z')R + 2x_n}\right),
\end{equation}

This is an equivalent result to that derived in \cite{SLAC-PUB-9242}, but for a qualitatively different reason; namely, the cancellation of terms between the velocity and radiation fields in the exit transient regime.

\section*{References}

\bibliographystyle{unsrt}
\bibliography{../references}

\end{document}